\documentclass[11pt]{arxiv-article}

\usepackage{bm}
\usepackage{needspace}
\usepackage{longtable}
\usepackage[table]{xcolor}
\numberwithin{equation}{section}

\DeclareFontFamily{U}{mathx}{}
\DeclareFontShape{U}{mathx}{m}{n}{
  <5> <6> <7> <8> <9> <10>
  <10.95> <12> <14.4> <17.28> <20.74> <24.88>
  mathx10
}{}
\DeclareSymbolFont{mathx}{U}{mathx}{m}{n}
\DeclareMathAccent{\widebar}{0}{mathx}{"73}

\newcommand{\prob}[1]{\mathfrak{#1}\xspace}
\newcommand{\order}[1]{\mathcal{O}\left(#1\right)}
\DeclarePairedDelimiterX\infdivx[2]{(}{)}{#1\;\delimsize\|\;#2}
\newcommand{\infdiv}[2]{\operatorname{D}_{\hspace{0.1mm}}\infdivx{#1}{#2}}
\newcommand{\mytuple}{y,t,x,s,\tilde s,z_1,z_2,z_3}
\DeclareMathOperator*{\argmin}{argmin}

\newcommand{\CECunit}{MW\text{ per Mio constant }\textsc{usd}}

\newcommand{\sectionlinetwo}[2]{%
  \nointerlineskip \vspace{.5\baselineskip}\hspace{\fill}
  {\color{#1}
    \resizebox{0.45\linewidth}{1.5ex}
    {{%
    {\begin{tikzpicture}
    \node  (C) at (0,0) {};
    \node (D) at (9,0) {};
    \path (C) to [ornament=#2] (D);
    \end{tikzpicture}}}}}%
    \hspace{\fill}
    \par\nointerlineskip \vspace{.5\baselineskip}
  }

\definecolor{bostonuniversityred}{rgb}{0.8, 0.0, 0.0}

\newcommand{\sidetitle}[2][right]{%
	\ifstrequal{#1}{left}{\reversemarginpar}{\normalmarginpar}%
	\marginnote{\raggedright\footnotesize\bfseries #2}%
}
\usepackage[textwidth=2cm,color=yellow]{todonotes}

\hypersetup{
	colorlinks=true,
	linkcolor=bostonuniversityred,
	citecolor=bostonuniversityred,
	urlcolor=bostonuniversityred
}

\usetikzlibrary{arrows.meta,calc,fit}

\definecolor{catA}{HTML}{70452F} 
\definecolor{catB}{HTML}{925D3B} 
\definecolor{catC}{HTML}{B17B50} 
\definecolor{catD}{HTML}{C5A66B} 
\definecolor{catE}{HTML}{B4BD79} 
\definecolor{catF}{HTML}{83A85D} 
\definecolor{catG}{HTML}{548A49} 
\definecolor{catH}{HTML}{286B3D} 

\newcommand{\entryspace}{\rule{0pt}{3.5ex}}   

\begin{document}	

\onehalfspacing

\thispagestyle{empty}

\begin{center}
{\Large {\bf \textcolor{bostonuniversityred}{
The Fiscal and Economic Implications of Green Mandates: \\[1ex]a Boltzmann Approach to Public Governance 
}
} }
\\[10pt]

\bigskip
{\large{\bf Katerina Savvaidou}}\hspace{0.02cm}\footnotemark[1] \,and\, {\large{\bf Georgia Foudoukidou-Louka}}\hspace{0.02cm}\footnotemark[2]
\\[2.5pt]
{\it Jean Monnet Chair on EU Fiscal Policy \& Administration\\
Faculty of Law, Aristotle University of Thessaloniki (AUTH)\\
University Campus, 54124 Greece}\\[1.5ex]  
and\\[1.5ex]
{\large{\bf Ho-Ryun Chung}}\hspace{0.02cm}\footnotemark[3]  
\,and\, {\large{\bf Orestis Loukas}}\hspace{0.02cm}\footnotemark[4]
\\[2.5pt]
{\it Institute for Medical Bioinformatics and Biostatistics\\
Philipps-Universität Marburg\\
Hans-Meerwein-Straße 6, 35032 Germany}

\bigskip
\end{center}

\footnotetext[1]{\href{mailto:savvaidou@law.auth.gr}{\texttt{savvaidou@law.auth.gr}}
	\qquad~~\,
	{\scriptsize $^{2}$} \href{mailto:georgiafl@law.auth.gr}{\texttt{georgiafl@law.auth.gr}}}

\footnotetext[3]{\href{mailto:ho.chung@uni-marburg.de}{\texttt{ho.chung@uni-marburg.de}}
	\qquad
	{\scriptsize $^{4}$} \href{mailto:orestis.loukas@uni-marburg.de}{\texttt{orestis.loukas@uni-marburg.de}}}

\begin{abstract}
	\normalsize
Public authorities are increasingly required to translate multifaceted policy objectives into concrete fiscal and investment choices while accounting for their socioeconomic consequences. In European economic and fiscal governance, for example, green budgeting provides a framework for translating environmental objectives into concrete targets for budgetary policy. Assessing the implications of meeting such targets before the implementation of concrete policy interventions calls for a credible ex-ante analysis.

The multiplicity of statistical models and data-processing pipelines raises the question of whether the historical baseline, reflected through a repository of infrastructure projects, and policy targets alone are sufficient to deliver a unique, tractable assessment.
To address this foundational question in the absence of more concrete knowledge about project selection, we bring the large-scale perspective of statistical physics, together with our earlier results on coarse-grained sampling probabilities, to public governance. 

For any policy objective expressible through a small number of aggregate conditions over many different projects, we construct, appealing to epistemic consistency under conditional sampling, the idealized policy scenario of minimal departure from the historical baseline. In Boltzmann's representation, this policy scenario takes the form of a counterfactual repository whose project composition exactly satisfies the specified policy targets while minimally departing from the historical repository. For a sufficiently large historical repository, so that details of project selection become effectively negligible, any dimension in this counterfactual repository is readily attributable to the historical baseline and specified policy targets.

Under the minimal-departure policy scenario, downstream implications of meeting the specified targets can be investigated based on any metric that was computable in the historical repository. Under the same policy scenario, the intrinsic $p$-value, constructed from coarse-graining the sampling probability of project repositories representing alternative historical versions, provides a statistical assessment of the targets' compatibility with the historical baseline.  To leading order in the number of historical projects, all downstream quantities can be efficiently estimated through closed-form expressions. 

As a proof-of-principle application, we analyze electricity-generation projects from the World Bank database. Our framework reveals a non-monotonic profile of expected project costs across feasible green-power surpluses, identifies the technological drivers of the green transition through their capital-efficiency coefficients, and exposes the risk of perpetuating existing regional disparities in funding and capital efficiency. Meanwhile, the historical green-power deficit remains statistically incompatible with any level of feasible green-power surplus.
\end{abstract}



\tableofcontents

\section{Introduction}

\subsection{Policy context and methodological requirements}

A climate mandate often requires a reallocation of various resources towards a specified policy objective. Policymakers seeking to accelerate a transition towards greener infrastructure need to assess not only whether the objective can be realized, but also what its implementation would imply for investment allocation and project costs, the technological mix, as well as further geographical and economic aspects. 
Within EU economic and fiscal governance, green budgeting provides a framework for integrating environmental objectives into budgetary decisions and assessing their broader fiscal and socioeconomic implications.	In particular, the fiscal implications of climate objectives can vary substantially with the policy instruments employed, with respect to the balance between public expenditure, private investment, subsidies, and revenue-side measures~\cite{Caselli2024}.
	
In fact, most policy objectives are inherently multidimensional.
In settings where policymakers have access to the historical baseline of an existing repository comprising public programmes with their expenditure or projects with their investments, a series of ex-ante questions arise.
Can a policy target be even realized with current technological capabilities and budget constraints? What would its implementation imply for capital reallocation across technologies and regions? What would happen to private investment and other dimensions that are not themselves under the policymaker's direct control? And, before proceeding with any restructuring, is the existing institutional and market environment, as reflected in the historical baseline, already plausibly compatible with the policy objective? Notice that such questions concern the implications of meeting a mandated target, rather than the estimation of an assumed underlying population model.

To answer such questions, the policymaker investigates the downstream consequences implied by a mandated target in an ex-ante impact assessment, which precedes and informs the strategic planning of an actual intervention pursuing this target.  
Such implications may include changes in fiscal expenditure, economic indicators, technological composition, regional distributions, or other relevant quantities. 
The policymaker can compare these implications with the applicable fiscal, economic, and policy constraints in order to assess whether the planned intervention  is compatible with the governing framework. To this end, principles of public governance, including proportionality and sound financial management, form part of the normative basis against which the magnitude and nature of downstream consequences are to be evaluated. 

For reported consequences to be meaningful and interpretable, their provenance must be clear from the assessment itself. {Established frameworks for evidence-based ex-ante impact assessment in EU and OECD regulatory governance emphasize that 
policy assessments should be grounded in clearly specified policy objectives, available evidence and an appropriate baseline, and should examine the consequences of alternative courses of action in relation to those objectives~\cite{ECBetterRegulation,OECD2012RegulatoryPolicyGovernance}.} Following these guidelines still does not determine a unique quantitative assessment, since
the same historical evidence and policy objective can support different analyses, depending on the choices made in their design. 
Such discretion creates scope for analysts, experts, and other stakeholders to substantially influence assessment outcomes and, consequently, the conclusions drawn from them~\cite{Hertin2009}.
More fundamentally, diverse assessments of the same policy objective raise the practical issue of selecting one estimate among several for the same metric; see e.g.\ a related concern in~\cite{GruneYanoff2018}. 

To facilitate pluralism in democratic deliberation, we need one quantitative answer to a clearly specified question --\,not to settle the debate, but to give it common ground from which meaningful disagreement and informed decisions can emerge.
After designating a \textit{historical project repository} from the available baseline, the ex-ante analysis of any policy objective should specify the mandated targets, relevant structural conditions, and, at most, a set of independently adopted principles 
relevant to public governance. 
An ex-ante impact assessment should be able to clearly isolate the contribution of these  core inputs to the reported consequences, disentangling it from choices made by analysts, experts, or other third parties.
We refer to this methodological requirement as \emph{attributability}. 

Next, we consider whether and how more conventional statistical approaches address attributability, before proposing an alternative statistical (re)formulation of the policy problem. Although we apply the proposed framework to the climate mandate of green transition, the underlying logic is not specific to climate policy. Our framework is readily generalizable to the ex-ante statistical assessment and downstream analysis of policy mandates in other areas of public governance.

\subsection{Conventional approaches to ex-ante assessment}
\label{ssc:ModelCentredApproaches}

Conventional quantitative approaches draw on statistical inference, predictive modelling, engineering models, or combinations thereof. Although these approaches can serve important scientific and predictive purposes, their design is not necessarily aligned with the ex-ante policy question considered here. 
The latter is concerned with the implications of meeting specified policy targets on the basis of an agreed historical repository. 
The assessment thus centres on attributability, rather than on inferring an underlying data-generating process or predicting future outcomes. Moreover, modeling the multitude of real-world influences relevant to such questions becomes not only computationally intractable, but also renders an exhaustive description of the problem epistemically impossible.

When model-centred approaches are nevertheless invoked for this task, they inevitably introduce simplifying assumptions, such as structural constraints and dependence relations that are not directly determined by the policy question itself or justified by some principle of public governance. Statistical modelling requires us to decide on the marginal and joint distributions of variables and the relationships between them, even when the task itself provides little incentive for such decisions. The literature on model uncertainty has long recognized the implications of modelling choices for policy conclusions~\cite{BrockDurlaufWest2003,Durlauf2012}. Established guidelines recommend sensitivity and uncertainty analyses, including comparisons across models, to investigate the implications of such choices~\cite{ECBetterRegulationToolbox,OECD2012RegulatoryPolicyGovernance}.

There seems to be no systematic, model-centred way to inherently eliminate the influence of additional constructs lying outside the scope of the policy question on the reported downstream consequences and statistical plausibility measures, such as the \(p\)-value. One might hope that their impact on the reported quantities remains small, but this cannot be intrinsically established within the ex-ante analysis itself. In a worst-case scenario, the influence of modelling could severely compromise attributability, with the reported consequences of a mandated target substantially affected by constructs foreign to the policymaker's objectives. 

For example, logistic regression is a standard method used across disciplines, often serving as a building block in more elaborate constructions. From a Maximum-Entropy perspective, the logistic model incorporates pairwise associations observed in the historical repository between the response and predictor variables, constraining higher-order effects to those implied by this pairwise structure. Adhering only to pairwise statistics from the historical baseline might be motivated by modelling stability or predictive accuracy. From the policymaker's perspective, it remains unclear though, which policy objective, structural condition, or governing principle would require this particular restriction. An attributability concern arises precisely because the modeled cubic and higher-order effects will generally differ from those present in the historical repository, even if the mandated targets were fulfilled in the historical repository. Crucially, the model provides no intrinsic way to establish whether the resulting discrepancies are negligible or substantial for downstream metrics entailing those higher-order effects. Hence, the policymaker is left unable to determine whether the reported consequences are indeed implied by the specified policy targets, as intended, or have been substantially altered by the model's structure. In spite of their transparent application, mathematically sound methods can therefore be seriously unsuitable for ex-ante assessments. 

Incidentally, the same concern underlies questions of \textsc{ai} explainability. Making an assessment's operation understandable does not automatically justify the stated assumptions on which its conclusions depend. Safeguarding attributability as a built-in requirement of ex-ante analysis motivates an alternative formulation that starts from the historical repository to incorporate the policy objective under explicitly specifiable constraints, without introducing ad hoc assumptions about the underlying processes.

\subsection{A statistical-physics perspective}
\label{ssc:StatisticalPhysicsPerspective}

The foundational modelling issue that pertains to most conventional statistical approaches, as described above, does not imply that ex-ante policy questions must remain beyond quantitative analysis. Given the complexity of repository formation, shaped by the interplay of social, economic, technological, physical, and institutional factors whose combined influence is difficult to disentangle, an alternative mathematical description is needed. This description should allow quantitative analysis without requiring an exhaustive account of the underlying mechanisms, while remaining grounded in those dimensions of the assessment that are directly specified by the policy question. 

One possible line of thought comes from the statistical physics of the 19th and 20th centuries. Statistical mechanics provides a systematic way to describe the collective behaviour of more elementary units and their large-scale organization, without requiring a detailed description of every microscopic state. Although such large-scale behaviour emerges from the collective contribution of many underlying mechanisms, its quantitative description, often mathematically tractable, can be formulated without specifying those mechanisms individually or explicitly modelling their interplay. The application of statistical-physics ideas to socioeconomic and policy problems has often taken the interacting agent as the elementary unit, using statistical-mechanical structures to describe how individual decisions and their interactions give rise to collective phenomena~\cite{Durlauf1996,Durlauf1999,Furtado2019}.

The perspective of statistical physics on socioeconomic problems indeed provides a valuable way of describing emergent regularities in collective phenomena~\cite{Castellano2009,DeMartinoMarsili2006}. As with any description of collective behaviour, its applicability depends on having a sufficiently large number of units. In our policy setting, a historical repository containing only a few projects would require detailed knowledge of its formation, including the decisions and circumstances that led to the projects' selection and implementation. If such information is unavailable, a generic statistical description cannot substitute for it. Introducing a more elaborate model not only does not automatically resolve the problem of missing information but also entails the danger of uncontrolled distortions.

The situation changes drastically when the historical repository contains sufficiently many projects for their selection to be treated as an approximately independent selection process. In this regime, the contribution of all real-world mechanisms to repository formation (understood e.g.\ through the picture of interacting agents or any other viable model of more elementary processes) can be collectively described without reconstructing or modelling each real-world process individually. With this perspective in mind, why not use the projects in a repository as the building blocks of the formal description? After all, they are what the policy question is usually about.

When the historical repository contains a large number \(N\) of projects, we shall \textit{effectively} treat the uncertainty caused by many real-world influences on project selection as ``luck'' or ``chance''. This is most easily demonstrated through a thought experiment (Gedankenexperiment) in which we repeatedly draw projects from an urn. The general relevance of this Gedankenexperiment, which describes what we would nowadays call conditional multinomial sampling, was first illustrated by L. Boltzmann in the later 19th century~\cite{boltzmann_weitere_2012}. Generic sampling from an urn thus conceptualizes the \textit{effective} statistical description of project selection motivated by collective behaviour in the appropriate large-\(N\) regime.\footnote{Uncertainty can be represented through different sampling schemes, such as hypergeometric or Poisson mixtures, which can lead to the same effective statistical description in appropriate limits. None of these conceptualizations is intended to reproduce what actually happened in the formation of the historical repository. They merely provide different ways of thinking about the same  statistical structure that universally emerges in the effective description.}

Remarkably, when the large-\(N\) conditions outlined above apply, this simplifying description, formulated directly in terms of distinct project profiles, can capture the emerging behaviour of aggregate quantities, such as average project cost, relevant to the policy question just as well as a description that explicitly accounts for all underlying mechanisms. 
According to the central intuition of statistical physics, when many influences contribute to an aggregate quantity, their details (which may well  explain how individual projects came about) become immaterial to the description of its  behaviour, provided no systematic mechanism prevents the large-\(N\) simplification. {If such a systematic distortion existed but remained unknown, any statistical description would be compromised.}

The conceptual simplification arising when many units collectively contribute to the quantities of interest does not automatically guarantee a computationally tractable evaluation of these quantities, even under an effectively multinomial sampling scheme (independent sampling from an infinite urn). Fortunately, policy questions are naturally expressed in terms of aggregate conditions on generalized moments, defined as expectations of functions of the physical, fiscal, or economic observables.
For precisely this class of questions that constrain generalized moments to prescribed values, our previous work~\cite{loukas2026boltzmannstructuresamplingintrinsic} has shown that the coarse-grained sampling probability mass, including the intrinsic \(p\)-value, admits a semi-analytic, closed-form expression to leading order in \(N\). For the derivation to remain valid, the number of aggregate conditions must be much smaller than the number of projects in the historical repository, which is in practice almost certainly the case.

Building on the tractability of the  effective description in terms of generic sampling, we use epistemic consistency of the intrinsic $p$-value and attributability to empirical structure to derive a least-departure counterfactual scenario.
The emerging $I$-projection that governs the saddle-point approximation to sampling probabilities describes the composition of an idealized, infinitely large project repository which minimally (in the information divergence sense) departs from the historical baseline to the extent necessary to satisfy the policy targets. In the illustrative Gedankenexperiment of the effective description, this counterfactual repository represents the idealized universe implied by the policy objective. 

Thus, exploiting the intrinsic structure emerging in the large-$N$ limit of generic sampling, we show --\,both formally and in practice\,-- that the policy question itself unambiguously constructs the counterfactual scenario of the ex-ante assessment, given the historical repository. By construction, these minimal ingredients automatically ensure attributability in the appropriate large-$N$ regime. This minimalism stands in contrast to the complexity and indirectness of model-centred approaches to the same problem that we outlined in Section~\ref{ssc:ModelCentredApproaches}.

Using the constructed counterfactual repository, we obtain both a statistical and a downstream ex-ante assessment. The statistical compatibility of the historical baseline with the mandated targets can be assessed, as can the mandate’s consequences through any metric computable from the historical project repository. 
The closed-form expressions (resulting from the Gaussianization around the saddle point), combined with the computational efficiency of conditional $I$-divergence minimization, make the framework suitable for repeated application as historical repositories are updated, policy objectives evolve, and new ex-ante assessments are undertaken. In fact, as the historical repository expands, the effective large-$N$ description is expected to become increasingly accurate. In contrast, model-centred approaches may require additional modelling of higher-order effects as the repository grows, adding to model complexity.

Beyond computational efficiency, the framework’s foundations in statistical physics, as opposed to an axiomatic statistical formulation, make it readily extensible to more general structural and policy conditions. Another natural path to explore are  generalized sampling schemes that incorporate procedural details from the actual project-repository formation, while retaining the guarantees of attributability.

\subsection*{Outline of the paper}

The paper proceeds from the designation of the historical repository of infrastructure projects to the formulation and assessment of counterfactual policy scenarios, exemplified in the context of green transition. First, we describe in Section~\ref{sc:DataPreparation} the data preparation and selection of sensible observables. These define the distinct project profiles that form the basic ingredients of the proposed Boltzmann framework for public governance.

Subsequently, we formulate the combined hypothesis corresponding to the green mandate, alongside intuitive structural conditions in Section~\ref{ssc:CombinedHypothesis}. 
Relying on the large-$N$ structure of coarse-grained sampling probabilities, the selection of the Boltzmann hypothesis distribution is justified in Section~\ref{ssc:BoltzmannDistro} and in greater detail in Appendix~\ref{app:Math}. 
The operational repercussions of the Boltzmann hypothesis counterfactual for statistical assessments in public governance are explored in Section~\ref{ssc:HypothesisTesting}.

Based on the World Bank database, Section~\ref{sc:Results} and Appendix~\ref{app:DownstreamAnalysis} present our ex-ante impact assessments of all feasible green-power surpluses through plots and figures. We examine the statistical compatibility of the mandated targets with the historical baseline and the implications of meeting these targets in the minimal-departure scenario across economic, technological, geographical, and investment-related dimensions. 

In Section~\ref{sc:Conclusions}, we conclude with the implications of our comprehensive ex-ante assessments for policymakers, including and beyond green budgeting. A broader outlook on extensions of the statistical-physics framework and its application to other policy settings is given in Section~\ref{sc:Outlook}. Table~\ref{app:BoltzmannFrameworkSummary} gives an overview of the interdisciplinary terminology used in the manuscript, while Figures~\ref{fig:FlowChart} and~\ref{fig:EffectiveProjectSelection} provide, respectively, an overall flow chart of the Boltzmann framework and a schematic depiction of project selection within its effective description.

\section{Historical data, project profiles and green classification}
\label{sc:DataPreparation}

The purpose of the following data preparation is to designate the historical project repository that serves as the empirical basis for our statistical analysis. The choices made in this section will influence the ex-ante assessment; hence, they must be very transparently documented. The Boltzmann framework does not attempt to eliminate the ambiguity arising from ``forking paths'' taken at this stage~\cite{GelmanLoken2013}. It neither optimizes nor otherwise corrects the designated project repository, which it accepts as an unquestionable input.

For a transparent demonstration, we use the World Bank Private Participation in Infrastructure (PPI) database~\cite{WorldBankPPI}. This choice is motivated by the availability of a substantial amount of publicly accessible, open-access data, rather than by any particular preference for the World Bank or its institutional and policy priorities.
While the raw database query can extend as far back as the beginning of the twentieth century, we restrict our analytical horizon to a forty-year observation window spanning from $1984$ to $2024$. This chronological cutoff is not arbitrarily selected, as it marks the historical kick-off of valid, structurally classifiable green infrastructure projects within global development pipelines. All projects in the query concern the electricity generation segment~\cite{WorldBankPPI_Glossary}.

\subsection{From raw data to observables and project profiles}
\label{ssc:Observables}

The primitive entities in the \textsc{ppi} database are distinct projects, up to certain redundancies that we filter out below. In our macroscopic description, we prioritize attributes that can be rigorously documented at the individual-project level over engineered administrative or econometric  indexes. 
Besides the requirement of \textit{observability}, namely that each attribute has a definite value for each project in the data, we impose no further constraint on the attributes we are going to analyze.
In particular, the subsequent analysis does not require us to postulate random variables or their underlying joint distributions.

We refer to such directly observed attributes selected for analysis as \emph{observables}. Having decided on a collection of observables such as $X,Y$, and $Z$, each project is represented by its corresponding $(x,y,z)$ tuple, uniquely identifying its relevant profile.  Throughout this work, we use uppercase Latin letters to denote the observables, while lowercase tuples denote the associated project profiles recorded in the given dataset. 

Trivially, if one were to employ as an observable some uninformative administrative marker such as \texttt{Project\,\,ID}, conditioning on it would make each project distinguishable, precluding any meaningful contribution to collective behaviour. By excluding such project-identifying labels from the policy question, profiles with common traits are aggregated together, with their empirical and counterfactual weights captured by the empirical and hypothesis distributions, respectively, as explained in Section~\ref{sc:Methodology}.

\paragraph{Physical observables.}
Each project profile includes the value $y$ of the project's nameplate \texttt{Physical\,\,Capacity} $Y$, measured in megawatts ($\mathrm{MW}$). 
To maintain demonstrative clarity, we keep preprocessing of the recorded values minimal. Projects whose $y$ entry is uninformative, such as \texttt{NaN} or \texttt{N/A}, are excluded rather than imputed. The physical capacity $Y$ is taken directly from the documented nameplate capacity at financial closure, without secondary adjustments for time-varying generation, seasonal effects, weather conditions, or other environmental factors. Such refinements can be introduced separately at a later stage without altering the core model-agnostic analysis.

Throughout the formulation of policy mandates, we retain physical observables in their original units rather than replacing them with engineered or derived metrics.
If a policy mandate were to require multiple observables, such as nameplate capacity together with rated frequency or energy storage capacity, all corresponding observables would simply be retained in the definition of the project profile, with their values recorded in $\mathrm{MW}$, $\mathrm{Hz}$, or $\mathrm{MWh}$, respectively. One could likewise supplement nameplate capacity with a count, such as the number of generating units, or with a categorical specification of the fuel or energy resource, without requiring numerical encoding.

Encoding heterogeneous observables via composite representations can introduce accidental dependencies, which may distort the subsequent statistical analysis in an uncontrollable fashion. Fortunately, the Boltzmann framework treats project profiles defined by any finite number of observables within the same formalism.  Instead of requiring observables to be reduced to a common test statistic through additional modeling choices, the intrinsic $p$-value (Section~\ref{ssc:HypothesisTesting}) is constructed directly from coarse-graining the sampling probability of alternative repositories containing those project profiles. In this way, the Boltzmann framework preserves the form of its statistical and downstream assessment irrespective of the number or type of observables defining the project profiles.

\paragraph{Economic observables.}
As our data span over four decades, there naturally exists a temporal observable $T$ representing the project vintage $t$, which we extract directly from the recorded \texttt{Investment\,\,Year}. Although $T$ carries the temporal direction of the project record, it is primarily an economic observable, since it records the year of investment rather than the physical age of the project.

The primary economic observable relevant to the fiscal-policy dimension of green budgeting is the \texttt{Total\,\,Investment} $X$, measured in millions of \textsc{us} dollars (\textsc{usd}). Since monetary values recorded over different vintages are expressed in nominal terms, we account for inflation across the observation window by converting each investment $x$ with vintage $t\in[1984,2024]$ into constant 2024 \textsc{usd} millions:
\begin{equation}
x_i \quad\rightarrow\quad c(t_i) \cdot x_i
\qquad \forall\, i\in\{1,\dots,N\}~,
\end{equation}
where the temporal adjustment factor $c(t)\in[1,\infty)$ is defined by the ratio of Consumer Price Indexes (\textsc{cpi}):
\begin{equation}
c(t)=\frac{\textsc{cpi}(2024)}{\textsc{cpi}(t)}~.
\end{equation}
Note this rescaling follows directly from the standard treatment of monetary values across different vintages in financial and economic analysis. Actually, this and similar transformations could be encoded directly at the level of conditions that control the fiscal aspect or via a downstream metric, leaving the original data intact. Beyond this transparently dictated adjustment, we apply no further transformations to the recorded values used in constructing project profiles.

\subsection{Green classification and policy observables}
\label{ssc:GreenClassification}

The implementation of a policy taxonomy --\,such as  France's 5-tier scale or the European Commission's framework dividing energy assets into three distinct categories of sustainable (green), harmful (brown), or transitional (neutral)\,-- is usually the product of an extensive preliminary analysis. This initial process usually incorporates both scientific and regulatory considerations through deliberation among technical experts, engineering researchers, institutional stakeholders, and policy practitioners. 

Our statistical framework operates downstream from this deliberation. It takes the adopted green taxonomy as a designated input, which is then encoded into the policy observable $S$, remaining agnostic to the political, regulatory or scientific criteria underlying this taxonomy. In contrast to an idea often propagated among scholars, there is no uncertainty in designated data. If a technology classified as green were later to be reclassified as environmentally harmful, the initial analysis should be considered wrong, requiring a re-evaluation using the updated classification.

The Boltzmann statistical framework can accommodate different modalities of $S$. A color classification scheme implies a categorical $S$, while percentages of a project's capacity tagged as green define a metric observable.
For clarity and ease of presentation in this initial study, we choose a binary tagging system based on projects' \texttt{Technology}: 
From the technology string $\xi$ recorded in each project's profile, we determine whether the listed power-generation technology can be consistently assigned to either side of the environmentally friendly/harmful divide, without ambiguity arising from multi-fuel technologies.
Subsequently, we assign to the profile the binary policy indicator $s$ according to
\begin{align}
    s =& \,\,\texttt{green\,\,renewables} \quad \text{if}\\
         \xi \in & \Big\{ \,
         \text{solar-PV}; \, 
         \text{solar-CSP}; \, 
         \text{solar-CPV}; \, 
         \text{wind}; \,
         \text{geothermal}; \,
        \nonumber\\
        & 
        \text{hydro-small (<50MW)}; \, 
        \text{hydro-large (>50MW)};\,
        \text{biogas}; \, 
        \text{biomass}; \, 
        \text{waste}; \, 
        \nonumber\\
        & 
        \text{solar-PV \& solar-PV};\,
        \text{solar-PV \& biogas}; \, 
        \text{solar-PV \& wind}; \, 
        \text{solar-PV \& not applicable};
          \nonumber\\
        & 
        \text{wind \& solar-PV}; \,
        \text{wind \& not applicable}  
         \Big\}\nonumber
\end{align}
and
\begin{align}
    s =& \,\, \texttt{brown/grey\,\,fossil}\quad \text{if}\hspace{9.5cm}\\
    \xi \in &\Big\{\, 
    \text{diesel}; \, 
    \text{coal}; \, 
    \text{natural gas}; \, 
    \text{steam}; 
    \nonumber\\
    &
     \text{diesel \& coal}; \,
    \text{diesel \& natural gas}; \, 
    \text{natural gas \& diesel}; \, 
    \text{natural gas \& steam}; 
     \nonumber\\
     & 
     \text{coal \& diesel}; \, 
     \text{coal \& natural gas}; \,
     \text{coal \& diesel \& natural gas};\,
      \nonumber\\
      &
     \text{not applicable \& diesel \& not applicable} \Big\}~. \nonumber
\end{align}
This simplified partition distinguishes between infrastructure projects based exclusively on low-carbon generation technologies and the conventional, carbon-intensive asset baseline.

To facilitate granular control over the green classification in our subsequent analysis, we map the raw text within the \texttt{Technology} attribute into an auxiliary policy observable $\tilde S$. This additional observable distinguishes between the specific green technologies collectively tagged as green. For each project profile, the value $\tilde s$ is determined by grouping similar green technologies according to:
\begin{equation}
\label{eq:GreenTechnologies}
    \tilde s = \left\lbrace
    \begin{array}{llr}
        \texttt{not\,\,applicable\,\,(N/A)} &\text{if}& s = \texttt{brown/grey\,fossil}\\
        \texttt{hydroelectric} && \text{hydro} \in\xi\\
        \texttt{solar\,\,fleet\,\,(PV/CPV/CSP)} && \text{solar} \in \xi\\
        \texttt{onshore\,\,wind} && \text{wind} \in \xi\\
        \texttt{geothermal\,\,heat} && \xi=\text{geothermal}\\
        \texttt{bio-energy\,\,\&\,\,waste\,\,(biomass/biogas/waste)} && \text{else}
    \end{array}\right.~,
\end{equation}
where $\xi$ denotes the raw string entry recorded under \texttt{Technology} and $s$ denotes the green tag defined in the previous paragraph.

At this point, we emphasize that the Boltzmann statistical framework could equivalently operate directly on the uncompressed \texttt{Technology} attribute and marginalize accordingly, without introducing primary and auxiliary policy observables.
The distinction between the primary policy observable $S$ and the granular policy observable $\tilde S$ was introduced here as a convenient way to present and discuss policy scenarios below. Clearly, the two observables are not independent sources of information, as both are deterministically derived from the same \texttt{Technology} attribute. 

Indeed, one could have equivalently defined a single policy observable comprising the five grouped green-technology categories together with one collective brown category. In that description, conditional sample means collectively referring to the green tag would have to aggregate explicitly over the five green categories when expressing the mandated conditions. The two-observable description separates the green/brown classification from the finer composition of green technologies, thereby enabling a clear formulation of mandated conditions in terms of the primary policy observable, while retaining the granular policy observable for the downstream analysis.

Finally, to ensure an unambiguous green classification, a strict exclusion filter is needed to prevent ambiguous assignments and eliminate multi-fuel contamination. Projects containing either controversial technologies (\texttt{Nuclear}) or green-brown hybrid combinations (e.g.\ \texttt{Hydro,\,\,Large\,\,(>50$MW$)} \& \text{Diesel} or \texttt{Natural\,\,Gas} \& \texttt{Steam} \& \texttt{Solar-PV})  are systematically excluded from the sample, alongside projects whose $\xi$ comprises uninformative metadata (\texttt{Not Applicable}, \texttt{Other}, and missing \texttt{NaN} variants).

\subsection{Secondary observables and complete project profiles}
\label{ssc:CompleteProjectProfiles}

In addition to the core $T,X,Y$ and policy observables $S,\tilde S$ introduced above, 
our study examines further observables denoted by $Z_i$ corresponding to the \texttt{Investment\,\,Share}, \texttt{Region}, and \texttt{IDA\,\,Status} attributes of the World Bank database.
These observables enter directly into the formulation of neither the green mandate nor the structural conditions in Section~\ref{ssc:CombinedHypothesis}. Nevertheless, the values of all secondary observables form part of the complete project profile and hence undergo the same strict data cleaning (no uninformative or corrupted values) as the primary observations. 

For later reference, the observables selected in the present study to define a complete project profile can be grouped into three categories. First, we have the {core observables} encompassing the physical observable $Y$, representing nameplate capacity and the fiscal-economic observables $X$ and $T$, representing investment amount and vintage. Second, we define the {policy observables} $S$ and $\tilde S$, representing the green--brown tagging and the granular green-technology classification. Finally, {secondary observables} $Z_i$ represent further economic and geographic properties relevant for an ex-ante analysis that assesses the downstream consequences of the green mandate. 

In the counterfactual project repository, the statistics of the selected secondary observables are jointly determined with those of the core and policy observables in a way that satisfies the green mandate alongside any exogenous constraints specified. The resulting joint statistics can subsequently be investigated in the downstream analysis. The above grouping is introduced here for bookkeeping and exposition only. It neither implies nor invites us to think about a corresponding decomposition of the effective description or any structured model for the joint behaviour of these observables. For the Boltzmann construction, what matters is the collection of observables that defines the project profiles and whether a policy hypothesis or structural constraint can be formulated quantitatively in terms of some or all of them. An observable need not enter one (or every) hypothesis, nor does its designation as core, policy, or secondary imply any methodologically different role in the counterfactual construction.

\subsection{The historical project repository}
\label{ssc:HistoricalRepository}

We conclude this section with the final form of the historical repository of infrastructure projects. Once agreed upon, the analysis shall not change it, i.e.\ there is effectively no residual uncertainty attributed to this historical baseline.

\paragraph{Identifying Distinct Projects.}
An empirical challenge when working with transnational project databases is the presence of duplicate contract logs, split financing tranches, and recurring transaction notes. Left uncorrected, these entries can cause overcounting of physical capacity and artificial inflation of the recorded investment.
In particular, the same asset can appear as multiple database entries across consecutive investment years with varying capital amounts, despite sharing an identical nameplate capacity. After standardizing the text within the \texttt{Project\:\:name} attribute across all entries in the database, we use this label to identify distinct projects: whenever the same project appears as more than one entry in the dataset, we first verify that the observed values match across key invariant observables, such as \texttt{Technology}, \texttt{Country}, \texttt{ContractPeriod}, etc. We then consolidate these entries into a unique project record. Its profile retains the invariant observed values, while the individual capital allocations are aggregated to obtain the total investment $x$, and the earliest recorded year is selected as the profile's vintage $t$.

Ultimately, this identification process reduces the number of entries in the dataset. 
More importantly, the identification of distinct projects is not only a matter of avoiding over-counting physical capacity. It also determines the actual number \(N\) of sampling units (projects) in the historical repository. If multiple database entries correspond to the same underlying project, treating them as separate observations would artificially increase \(N\), thus introducing unjustified sensitivity into hypothesis testing, besides altering the empirical frequencies of project profiles.
This $N$ also sets the upper bound on the number of distinct project profiles observable in the dataset.

\paragraph{The resulting sample and acceptance threshold.}
Applying the preceding de-duplication and filtering protocols without introducing arbitrary data imputations or synthetic variables, we obtain a dataset that 
maximally preserves the raw empirical statistics of the custom query without further modification. The resulting sampling size of
\begin{equation}
    N = 3\,729 \text{ unique infrastructure data points}
\end{equation}
motivates the conservative acceptance  threshold (the so-called significance level in conventional hypothesis testing)
\begin{equation}
    \alpha = \frac{1}{N} = 0.000268 \approx 0.03 \% \ll 5\%~.
\end{equation}
This threshold provides a stringent decision boundary for the hypothesis testing of counterfactual scenarios in Section~\ref{ssc:DesicionTests}, such that statistical rejections require evidence substantially beyond the level associated with a single sampling unit (project).

To avoid common misinterpretations, we emphasize that we do not bin, smooth, or otherwise process the continuous metric observables such as physical capacity $Y$, capital allocation $X$, or temporal observables such as project vintage $T$. The recategorization applied to the \texttt{Technology} attribute serves solely to encode an adopted green classification through the binary policy observable $S$ and its granular counterpart $\tilde S$. This classification is an exogenous policy choice, not dictated by any mathematical restriction of the statistical framework which could also handle metric policy observables. The categories assigned to $S$ could instead be defined according to any other taxonomy agreed upon through the relevant policy process, without altering the implemented formulation.

\section{The counterfactual world of Boltzmann sampling}
\label{sc:Methodology}

With our transparently processed and complete dataset designating the historical repository, we now turn to the construction of counterfactual policy scenarios. For policymakers, such counterfactuals provide idealized policy universes in which the potential consequences of a policy objective can be examined without requiring the corresponding intervention to be actually implemented in the infrastructure.

To exploit the anticipated large-$N$ simplification outlined in Section~\ref{ssc:StatisticalPhysicsPerspective}, we employ the combinatorial framework developed in~\cite{loukas2024totallyempiricalbasisscience,loukas2026boltzmannstructuresamplingintrinsic}. Rather than presupposing an abstract probability space or an underlying population distribution, our approach starts from the finite collection of observed projects and constructs a counterfactual repository. 
In the policy setting, we refer to a counterfactual repository that exactly realizes the intended policy objective as a \emph{hypothesis counterfactual}. In this idealized repository, thought of as an infinitely large urn, the projects themselves serve as the sampling units (or balls in the urn), distinguishable up to their profiles (the types of balls), which are defined by the observables selected for the policy question and intended downstream analysis. Different hypothesis counterfactuals correspond to different policy universes 
in which the unknown mechanisms of project selection admit the effective description of generic sampling.

In the present study, the complete project profile is dictated by the core, policy and secondary observables $Y$, $T,X$, $S,\tilde S$ and $Z_1,Z_2,Z_3$ respectively, as selected in Section~\ref{sc:DataPreparation}. In that way, each distinct  profile $(\mytuple)$ identifies a type of ball in the urn. 
A hypothesis distribution $\prob p$ assigns a weight to each --\,in principle\,-- observable profile indexed by this 8-tuple, thereby specifying the composition of a hypothesis counterfactual through reweighting alone. In the large-\(N\) regime, an alternative repository of \(N\) projects drawn from this hypothesis counterfactual can be treated, at least for the purposes of the intended impact analysis, as the outcome of generic sampling, in which projects are drawn independently.

The non-axiomatic framework does not require observables to be declared a priori as metric or nominal. The modality with which an observable enters the hypothesis logically follows from the conditions imposed on it: a condition on a mean or variance naturally gives it a metric modality, while a marginal prevalence gives it a categorical modality. Conversely, a condition may concern particular values or categories of a metric observable. In the Boltzmann construction, these distinctions do not affect the effective reweighting mechanism, which is indifferent to them. This leaves the modality of an observable as a question of how the concrete policy mandate is to be formulated as a condition involving that observable.

Two further features of our statistical-physics perspective are particularly important for the present application with minimal knowledge about the underlying project selection process. Coarse-graining the sampling probability within the universality class of generic sampling provides an intrinsic \(p\)-value, without assuming a population model or introducing an application-specific test statistic whose reference distribution must be established asymptotically. Having derived semi-analytic closed formulas that robustly compute at large sample size $N$ this \(p\)-value, we no longer need to rely on computationally intensive, brute-force bootstrapping to generate synthetic datasets. 
In a unifying manner, the hypothesis distribution constructed for each counterfactual policy scenario, which enables its downstream analysis, also directly enters the test statistic determining the intrinsic $p$-value.

\paragraph{The empirical standpoint.}
The empirical distribution, denoted by $\prob f$, assigns relative frequencies to the distinct project profiles observed in the historical repository designated in Section~\ref{ssc:HistoricalRepository}. These profiles define the ball types in the urn of our Gedankenexperiment. Notice that the ``elementary'' within the description property of a project, its profile, depends entirely on the assessment at hand. Different policy mandates and downstream metrics generally call for different observables and hence different project profiles, as discussed in Section~\ref{ssc:CompleteProjectProfiles}.

Unless all possible combinations of the values individually assumed by the selected observables were observed in the historical repository, the set of observed project profiles would not nicely factorize into the individual observables’ value sets. In other words, the set of observed project profiles would not exhaust the Cartesian product of the individual value sets, let alone exhibiting some product structure among the mathematical domains. This happens due to either finite sample size or structural constraints, potentially unknown to the analyst, that make some combinations technically or administratively non-admissible. For example, a modular green-power installation with a reasonable budget cannot have the nameplate capacity of a large hydroelectric plant.

As a consequence of the decision not to extrapolate onto unseen profiles that would require a more detailed understanding, the empirical distribution \(\prob f\) contains strictly positive weights. 
More intriguingly, the non-product structure of the set of profiles hinders the derivation of closed-form expressions for the hypothesis distribution in terms of the prescribed generalized moments, even under simpler linear conditions. In model-centred (see also Maximum-Entropy) approaches, this might at first seem like an undesirable feature.
However, it is precisely this non-trivial empirical structure, as we argue in Section~\ref{ssc:BoltzmannDistro}, that leads to a well-defined answer to the ex-ante policy question readily ensuring attributability. 
In practice, the lack of a closed-form expression for the hypothesis distribution in terms of controllable parameters has no effect on the construction of hypothesis counterfactuals based on the $I$-projection (Section~\ref{ssc:BoltzmannDistro}) and on their statistical assessment which exploits the large-$N$ semi-analytic closed-form expression for the $p$-value (Section~\ref{ssc:HypothesisTesting}).

\subsection{The hypothesis of a green mandate}
\label{ssc:CombinedHypothesis}

While our framework can readily accommodate more complex policy mandates involving multiple target observables $Y_1, Y_2,\ldots$, we deliberately focus here on a simpler baseline demonstration in terms of nameplate capacity $Y$ discussed in Section~\ref{ssc:Observables}. In what follows, we take the policy mandate to be that the overall nameplate capacity tagged as green exceeds the nameplate capacity tagged as brown in Section~\ref{ssc:GreenClassification}.

Specifically, we evaluate the effects of enforcing power balance $\eta=0$, as well as varying target levels of green-power surplus $\eta>0$ or deficit $\eta<0$ (for retrospective consistency), which are represented by the expected power difference, 
\begin{equation}
\label{eq:EtaCondition}
\eta \overset{!}{=} \sum _{y}y\cdot p(y,\texttt{green\,renewables})-\sum _{y}y\cdot p(y,\texttt{brown/grey\,fossil})
~.
\end{equation}
Here, the hypothesis distribution $\prob p$ describes the composition of a counterfactual repository, assigning a specific probability $p(\mytuple)$, i.e.\ a non-negative weight, to each observed project profile $(\mytuple)$. For notational simplicity, any shorthand indexing such as $p(y,s)$ above, which uses only a subset of the observables, is understood to represent the corresponding marginal of $\prob p$, obtained by summing over all jointly observed values of the remaining observables.

Using the empirical distribution $\prob p=\prob f$ --\,intuitively understood as the multi-dimensional histogram of our designated historical repository, we directly obtain  the difference of sample means in Eq.~\eqref{eq:EtaCondition}, 
\begin{equation}
	\label{eq:GreenPowerDeficit}
	\mu = \sum _{y}y\cdot f(y,\texttt{green\,renewables})-\sum _{y}y\cdot f(y,\texttt{brown/grey\,fossil})~.
\end{equation} 
A historical green-power deficit associated with a negative difference fails to comply with forward-looking policy objectives that target a green-power surplus, i.e.\ a positive power difference. Even a positive difference in the historical baseline does not automatically imply compliance with the mandated target. This mismatch in sign or magnitude motivates the construction of counterfactual repositories. Conversely, if the historical baseline already met the mandated target(s), there would be no need to construct some ``unbiased'' version of the historical repository, which remains the only designated factual baseline. As we shall see in Section~\ref{ssc:HypothesisTesting}, the intrinsic \(p\)-value would in that case become 1.

\paragraph{Structural conditions.}
In our demonstrative scenario, we need to hold specific aggregate physical and financial quantities constant to reflect real-world operational and institutional constraints. Within the Boltzmann framework, these exogenous constraints are formulated as structural conditions on admissible hypothesis counterfactuals. For instance, a regulatory authority might evaluate policy shifts while locking in a total capital budget or a technology-specific financial allocation.

As an intuitive paradigm, we evaluate the transition toward a green power surplus while holding the total nameplate capacity across all regions and project vintages fixed. 
Mathematically, we fix the first moment of observable $Y$, since $N$ times its sample mean determines the aggregate capacity in the historical repository. While this first moment can theoretically be calibrated to any physically sensible value supported by the observed profiles, we set it to its empirical value due to a lack of prior administrative directives:
\begin{equation}
	\label{eq:TotalCapacityCondition}
	\sum_y y \cdot p(y) \overset{!}{=} \sum_y y \cdot f(y) ~.
\end{equation}

The global condition on the first moment of $Y$ can be interpreted as a basic physical constraint on the delivery of additional electricity-generation capacity. In the present setting, $Y$ represents the additional nameplate capacity historically committed through the World Bank projects. Maintaining its overall historical level ensures that the policy mandate does not, by itself, imply an aggregate increase or reduction in the additional nameplate capacity relative to the historical commitment. From an engineering perspective, deviations in either direction could contribute to generation curtailment or supply shortfalls, respectively, depending on the capacity already available in the electricity system. 

Obviously, a detailed analysis of the actual worldwide electricity system would require additional information on its physical structure and operation, including regional demand, dispatch, storage, transmission and interconnection capacities, and electricity flows between regions. Such an analysis is beyond the scope of the present demonstrative example and cannot be performed from the historical repository used here alone. In particular, regional capacity constraints could be incorporated as additional structural conditions when the corresponding information and policy requirements are available.

Since the hypothesis counterfactual is defined through a probability distribution specifying the composition of a not necessarily finite repository, it is challenging to interpret a rescaled mean difference in Eq.~\eqref{eq:EtaCondition} as an overall green-to-brown power difference. Once total capacity is fixed, however, the corresponding power difference as a share of total capacity provides a well-defined dimensionless measure of the mandated green-power surplus or deficit. This normalized quantity is used as the controlling parameter throughout the plots of Section~\ref{sc:Results}. 
Accordingly, the green-to-brown power difference and the green-to-brown power ratio, whenever the latter is well defined, become equivalent ways of parametrizing the green mandate at fixed total capacity. In particular, the ratio can be expressed solely in terms of the targeted power difference in Eq.~\eqref{eq:EtaCondition} and the fixed total capacity in Eq.~\eqref{eq:TotalCapacityCondition}.

Furthermore, we introduce in our counterfactuals an operational boundary condition. Invoking the granular policy observable $\tilde S$, we hold hydroelectric capacity constant across the green transition through the strictly invariant moment
\begin{equation}
\label{eq:HydroCondition}
\sum _{y}y\cdot p(y,\texttt{hydroelectric}) \overset{!}{=} \sum _{y}y\cdot f(y,\texttt{hydroelectric})~.
\end{equation}
The physical, geographical and environmental aspects of hydroelectric generation motivate this exogenous constraint. Unlike modular renewable technologies, hydroelectric capacity cannot be scaled arbitrarily in response to a policy mandate. A nation cannot simply engineer high-altitude rivers or lakes to build massive dams. In fact, accelerating climate change and recurring droughts directly threaten existing water reservoirs, thereby also the hydroelectric baselines. In light of these concerns, we do not allow a counterfactual green transition to be driven by an aggregate hydroelectric commitment above its already vulnerable historical level.

\paragraph{The combined hypothesis.}
The mandated target formulated in Eq.~\eqref{eq:EtaCondition} together with the structural conditions Eq.~\eqref{eq:TotalCapacityCondition} and Eq.~\eqref{eq:HydroCondition} defines the combined hypothesis of the policy objective. 
In that way, the combined hypothesis specifies a linear family of distributions over the observed project profiles that satisfy the green mandate and all imposed structural conditions. Any distribution within this linear hypothesis family unambiguously describes a hypothesis counterfactual that satisfies the combined hypothesis.

In the large-$N$ regime, which is achievable with the extensive repositories maintained by international organizations such as the World Bank, the number of distinct project profiles can also grow with $N$. Such concurrent scaling reflects the multidimensional nature of most policy questions which aim at analyzing multiple observables downstream. Conversely, a historical repository containing only e.g.\ three projects and subject to two independent conditions is too small to warrant a statistical assessment in the first place. In such a setting, attributability could hardly be guaranteed without introducing additional assumptions.

Most policy questions are formulated across only a much smaller number of dimensions, translating into a handful of independent mandated and structural conditions defining the combined hypothesis. (In our demonstrative application, we have stated four conditions, one mandated, and two non-trivial structural conditions in addition to normalization on the simplex). This is precisely the large-$N$ regime we are interested in, whose tractable statistical structure has been entirely overlooked in the literature. Irrespective of a concurrent growth in the number of project profiles, potentially reaching \(N\) distinct profiles in the extreme case, which marks the breakdown of classical Sanov theory, the coarse-grained sampling probabilities still admit semi-analytic closed-form expressions~\cite{loukas2026boltzmannstructuresamplingintrinsic}.  

A combined hypothesis that defines an under-specified linear system leaves residual freedom regarding the composition of hypothesis counterfactuals. When the linear system admits at least one distribution, the combined hypothesis is said to be \textit{feasible} within the range of historical project profiles. This signals that the policy target can be achieved without invoking unobserved project profiles beyond the historical baseline.  In the under-specified regime, the linear family of a feasible combined hypothesis contains infinitely many distributions.

\subsection{The Boltzmann hypothesis  distribution}
\label{ssc:BoltzmannDistro}

The combined hypothesis specifies the conditions that a hypothesis counterfactual must satisfy, but it does not determine a unique hypothesis counterfactual in the under-specified setting commonly encountered. 
This degeneracy reveals that the conditions following from the mandated policy alone, even when supplemented by exogenous constraints, do not automatically lead to the hypothesis counterfactual, leaving multiple downstream answers possible.

Fortunately, interpretability of the intrinsic $p$-value motivates the forward $I$-projection of the empirical distribution $\prob f$ onto the linear family defined by the combined hypothesis as the hypothesis distribution. This projection is unique whenever the hypothesis linear family is non-empty. For each feasible mandate, we thus provide a principled selection criterion that determines a single hypothesis distribution, the so-called \emph{Boltzmann hypothesis  distribution}, which in turn specifies the composition of the hypothesis counterfactual. 
In particular, this \textit{Boltzmann hypothesis  counterfactual} is used in the downstream analysis as a minimally disruptive scenario, with the dimensions left unspecified by the combined hypothesis unambiguously determined by the \(I\)-projection.

In the following paragraphs, we give the core epistemic argument for the selection of a hypothesis distribution. After outlining the meaning of the intrinsic \(p\)-value through our working example in public governance, we state the condition for its interpretability in statistical hypothesis testing and explain how it naturally leads at large \(N\) to the Boltzmann hypothesis  distribution. Ultimately, we show that sampling conditional on the historical green-power deficit under the Boltzmann hypothesis counterfactual is dictated by the additional empirical structure of the historical repository.
This empirical attributability of the additional structure in conditional sampling is a property that singles out the Boltzmann  hypothesis counterfactual. 

\paragraph{Generic sampling.}
In the large-$N$ regime, the collective influence of many mechanisms contributing to repository formation can be described, for the purposes of the present ex-ante analysis, through generic sampling, in which projects are drawn independently from a hypothesis counterfactual. In the language of statistical physics, the regime in which the details of a system, or more generally a process, become immaterial to the description of investigated quantities within an acceptable approximation is called \textit{universality}. Universality is not a property conferred by an externally selected model, but a feature of the real-world application itself in the relevant limit. Generic sampling provides a minimal description through the multinomial scheme without requiring the individual mechanisms contributing to repository formation to be specified.

Extending Boltzmann's original Gedankenexperiment provides a concrete, physical realization of the generic sampling process. Specifically, we consider \emph{alternative repositories} of the same size \(N\) as the historical repository, drawn from a hypothesis counterfactual and satisfying the structural conditions Eq.~\eqref{eq:TotalCapacityCondition} and Eq.~\eqref{eq:HydroCondition}. Such conditional sampling of alternative repositories can be obtained through rejection sampling from the corresponding hypothesis distribution. These finite project repositories represent alternative historical versions generated under the selected hypothesis counterfactual.

Due to sampling uncertainty --\,representing in the urn scheme the collective effect of real-world influences\,-- the aggregate power difference Eq.~\eqref{eq:EtaCondition} will differ among alternative repositories.  
Based on this aggregate power difference, it is thus natural to group  alternative repositories into \textit{repository families}. By construction, all alternative repositories belonging to a family support the same mandated $\eta$ in  Eq.~\eqref{eq:EtaCondition}, in addition to sharing the same empirical estimates  of the generalized moments in Eqs.~\eqref{eq:TotalCapacityCondition} and~\eqref{eq:HydroCondition}.
One such family exhibits the historical green-power deficit. By definition, the historical repository itself belongs to the \emph{historical repository family}.

\paragraph{The emergence of the information projection.}
By coarse-graining the probability over alternative repositories in the conditional sampling scheme, each repository family is assigned a probability mass that depends on the chosen hypothesis counterfactual. Provided the large-\(N\) conditions apply, 
the saddle-point approximation to this probability mass determines the leading-order conclusions for the ex-ante assessment, with the effects of a more detailed description of repository formation entering at subleading orders.
The intrinsic $p$-value 
is the cumulative probability of sampling any repository family from the hypothesis counterfactual  at least as improbable as the historical repository family corresponding to the observed green-power deficit in Eq.~\eqref{eq:GreenPowerDeficit}.

Whenever $N$ is sufficiently large relative to the number of hypothesis conditions, conclusions can be universally drawn from this intrinsic \(p\)-value, which admits a $\chi^2$ approximation. As we derived in~\cite{loukas2026boltzmannstructuresamplingintrinsic}, the emerging test statistic is given by twice $N$ times the $I$-projection. 
In the current application, involving one mandated target, the intrinsic \(p\)-value is approximated using the cumulative distribution function of a \(\chi^2\) distribution with one degree of freedom. The $I$-projection is obtained by minimizing the information divergence from the chosen hypothesis distribution over the historical repository family, since we have restricted our analysis to hypothesis distributions that reproduce the empirical estimates of the structural moments.

\paragraph{The epistemic requirement.}
Having established this level of tractability for an otherwise highly complex real-world problem without relying on external model assumptions,  we must settle the hypothesis distribution that enters the computation of  the \(p\)-value. To do so, we consider the operational use of the intrinsic $p$-value for the statistical assessment of an ex-ante policy objective. The epistemic considerations that follow evidently apply to any statistical assessment using a sampling-based \(p\)-value, independently of the particular application.

For the intrinsic \(p\)-value to support a meaningful decision against a prescribed acceptance threshold, the historical repository must remain the most probable alternative repository among all those that support the observed green-power deficit. 
Otherwise, a high \(p\)-value could  
paradoxically suggest that the observed  deficit were sufficiently probable under the hypothesis counterfactual, even though the historical repository from which it had been obtained would itself appear, especially at larger $N$, highly improbable. 
Such a paradoxical interpretation must be prevented, since the designated historical repository is, after all, the only invariant in the ex-ante assessment.

We adopt this requirement as an epistemic criterion for the interpretation of the intrinsic \(p\)-value.
At large $N$, the aforementioned paradoxical situations can be ruled out by requiring the historical repository to be the most probable member of the historical repository family.
As shown in Appendix~\ref{app:Math}, this requirement of Maximum Probability is fulfilled by the Boltzmann hypothesis  distribution as the forward $I$-projection of the empirical distribution $\prob f$ onto the linear family defined by the combined hypothesis.  
Provided the combined hypothesis is feasible, the relevant $I$-projection is unique~\cite{csiszar_i-divergence_1975}, thereby offering a principled mechanism to select  the hypothesis counterfactual. 

Had we incorporated more detailed information on the formation of project repositories, leading to a different sampling scheme, such as a generalized Pólya urn that accounts for finite-population effects in the hypothesis counterfactual, another divergence measure would have emerged in the appropriate limit of physical scales (see Appendix~A of~\cite{loukas2026boltzmannstructuresamplingintrinsic}). Ultimately, this would determine another hypothesis distribution under the same epistemic criterion. For an attributable ex-ante impact assessment, the dependence on the emerging sampling scheme makes it imperative to state transparently which details of repository formation are incorporated, since no epistemic principle is going to prescribe those. 

\paragraph{An auxiliary picture.}
In the present setting of linear conditions on generalized moments and effectively multinomial sampling at large $N$, the Boltzmann hypothesis  distribution selected by the epistemic criterion can be understood also through a conditional multinomial sampling scheme. We repeatedly throw an unfair multifaceted die, whose face probabilities are given by the historical relative frequencies of observed project profiles, and record the resulting infinitely large counterfactual repository only if it satisfies the combined hypothesis. 
Among these admissible counterfactuals, the repository whose composition is dictated by the Boltzmann hypothesis distribution is the only one surviving with nonzero probability in the idealized infinite-size limit.
This auxiliary picture follows from a classical Sanov-type expansion of the conditional sampling probabilities with the empirical distribution \(\prob f\) now taken as the reference.
Non-linear conditions on generalized moments or other sampling schemes need not admit such a sampling-based understanding of the consequences of the epistemic criterion.

\paragraph{Attributability of counterfactual generalized moments to historical baseline.}
In realistic applications, there are many more dimensions in the repository not directly specified by the combined hypothesis of Section~\ref{ssc:CombinedHypothesis}. Such additional dimensions can be captured by collections of generalized moments. A hypothesis counterfactual prescribes concrete values for them which in turn influence downstream metrics as well as the intrinsic $p$-value. Unless the mandate is exactly fulfilled in the historical repository, the counterfactual values will differ from the historical baseline. 

Among possible hypothesis counterfactuals, only the one whose composition is dictated by the Boltzmann hypothesis distribution prescribes values for all additional generalized moments which are fully attributable to their historical estimates under conditional sampling. This time, the conditional scheme involves coarse-graining the sampling probability of alternative repository families that are nested within the historical repository family. Only under the Boltzmann hypothesis counterfactual does the intrinsic \(p\)-value to sample the historical estimate or an at least as improbable value for \textit{any} additional generalized moment, given the historical green-power deficit, equal one. 
Such nested assessments should not be conflated with the statistical assessment of the historical green-power deficit itself, which assesses the deficit under the structural conditions.

Specifically, we define a nested repository family by 
constraining further generalized moment(s) to their historical estimates, in addition to the structural conditions Eqs.~\eqref{eq:TotalCapacityCondition},~\eqref{eq:HydroCondition} and the condition on the historical green-power deficit Eq.~\eqref{eq:GreenPowerDeficit}. 
Under the same hypothesis counterfactual, we assess statistically the nested repository family relative to the ambient historical repository family. Attributability to the historical baseline means that the intrinsic $p$-value of this statistical assessment must always remain one (in saddle-point approximation), no matter which additional generalized moment is used.

The epistemic argument for the interpretability of the intrinsic $p$-value introduced above can also be understood as the extreme case of attributability to the combined additional structure of the historical repository. For that, we note that the historical repository is the only member of the nested family defined by 
specifying the historical estimates of the 
remaining dimensions that complement the structural and mandate conditions over the finite set of observed project profiles.
The $p$-value for this maximally specified nested family 
is automatically one under the Boltzmann hypothesis counterfactual, where the historical repository has the highest sampling probability  within the ambient historical repository family. 

\paragraph{Linear-algebraic formulation and numerical computation.}
Since the number of observed project profiles is bounded by $N$, a distribution $\prob p$ can be represented by a column vector containing the probabilities of distinct profiles. Accordingly, a condition like Eq.~\eqref{eq:EtaCondition} that prescribes sample means in the counterfactual described by $\prob p$ can be realized via the scalar product of such a probability column vector with a row vector whose coefficients equal either $y$ for \texttt{green\,\,renewables} or $-y$ for \texttt{brown/grey\,\,fossil} profiles. To compute the prevalence of \texttt{hydro-electric} technology, for instance, we would instead use  a row vector of ones for project profiles with $\tilde s=\texttt{hydro-electric}$ and zeros otherwise. The normalization condition is given by the constant row vector of ones for all project profiles. 

To summarize a collection of conditions, we can use the coefficient matrix  whose rows consist of the aforementioned row vectors.  Acting with this coefficient matrix on a probability column vector  produces another column vector containing the  values of the corresponding generalized moments. 
The outlined linear-algebraic formulation~\cite{loukas2024totallyempiricalbasisscience} provides a convenient way to organize the conditional information-divergence minimization. In particular, after filtering out redundant conditions, we can robustly identify the relevant degrees of freedom that enter the subsequent hypothesis testing. In the current setting, these degrees of freedom are obtained by subtracting the row rank of the coefficient matrix comprising the structural conditions from the row rank of the coefficient matrix of the combined hypothesis.

In the absence of project profiles whose weight is forced to zero by the combined hypothesis, the Boltzmann hypothesis distribution belongs to the exponential family generated by the coefficient matrix, with $\prob f$ as reference. This familiar mathematical form is not an external assumption imposed on the counterfactual. On the contrary, it emerges from insisting on a consistently interpretable intrinsic \(p\)-value for our statistical assessments.
Incidentally, the resulting least-bias selection with respect to the original data recovers Efron's exponential tilting~\cite{efron_bootstrap_1979} in the bootstrap setting, albeit without the need for downstream bootstrap. 

Exploiting its exponential form, we can parameterize the Boltzmann hypothesis distribution in terms of the independent conditions after filtering out linearly dependent ones. The corresponding profile probabilities can then be numerically determined using the Newton-Raphson root-finding algorithm.\footnote{For an optimized \texttt{C++} implementation of the $I$-projector with \texttt{Python}/\texttt{R} wrappers, see \href{https://gitlab.uni-marburg.de/fb20/ag-chung/iprojector/}{https://gitlab.uni-marburg.de/fb20/ag-chung/iprojector/}.} This linear-algebraic parametrization streamlines the estimation of the forward \(I\)-projection through a fast and robust numerical procedure that determines the non-trivial minimum of the information divergence under linear conditions.

\subsection{Operational implications}
\label{ssc:HypothesisTesting}

For a specific green-power surplus, there are in principle many hypothesis counterfactuals composed of observed project profiles, all compatible with the designated structural conditions. These would differ in their downstream metrics, such as the investment allocation or capital efficiency coefficients. This plurality of answers makes it necessary to establish a unique reference point for democratic deliberation and decision-making.

Having the Boltzmann hypothesis distribution provides a central payoff of the proposed paradigm shift. Without concretely specifying anything about the actual selection of projects, the composition of the Boltzmann hypothesis counterfactual is, by construction, attributable to the historical baseline and the combined hypothesis comprising the structural conditions and mandated target.
In particular, this attributability guarantees a compositional aspect of \emph{additionality}. In public-policy evaluation, additionality has commonly been discussed in connection with policy attribution~\cite{OECD2022SoilCarbon,OECD2024Additionality}.  In the Boltzmann framework, the absence of a policy objective means no departure from the historical repository, see also the paragraph on the empirical standpoint in the introductory part of Section~\ref{sc:Methodology}.
That the counterfactual departure from the historical repository is solely triggered by the mandated target under the prescribed setting, does not imply or assert, though, a causal counterfactual about what would have happened in the absence of the policy objective. 

\paragraph{Alternative historical versions in an idealized policy universe.}
The Boltzmann hypothesis counterfactual describes the policy universe envisaged by the policymaker. In this counterfactual repository that minimally departs from the historical repository, the project composition satisfies the mandated green-power surplus while respecting any designated structural conditions. (We emphasize that a hypothesis counterfactual does not represent a prediction from the historical baseline.)
Within this idealized universe, had the policy target been adopted, the policymaker asks what would happen when the adopted policy met the many influences that shape real-world project repositories. 
Administrative inertia, fragmented country-specific policies, socio-economic developments, political decisions, engineering constraints, and climate fluctuations all affect which projects materialize and in what proportions. 

Since the ways that this happens cannot be fully anticipated or explicitly described in great detail, finite repositories that could materialize within a counterfactual universe, representing alternative historical versions, would not exactly adhere to the composition of the hypothesis counterfactual. For example, the first moment of the fiscal observable $X$, representing the project-cost estimate in a given alternative repository, would generally differ from its anticipated estimate in the hypothesis counterfactual.  Nevertheless, the \textit{expectation} of a generalized moment\footnote{In the conditional sampling of alternative repositories, the expectation of a generalized moment is expressed as a sum over its feasible values, weighted by the coarse-grained sampling probabilities of the corresponding repository families, defined by the moment's values and structural conditions. This sum can be approximated using large-$N$ saddle-point techniques via a lower-dimensional Gaussian for any number of historical project profiles.},  obtained by averaging its different values across alternative repositories, will reproduce  its counterfactual estimate in the idealized policy universe given by the Boltzmann hypothesis distribution.

Following the statistical-physics perspective on collective phenomena, we represent the uncertainty arising from unresolved influences through generic sampling of alternative repositories of the same size $N$ as the historical repository. 
In the urn Gedankenexperiment,
one thinks of projects as balls of different types corresponding to their distinct profiles. In this picture (Figure~\ref{fig:EffectiveProjectSelection}), an alternative repository is formed by drawing $N$ balls from an infinite urn whose composition (prevalence of types) is given by the Boltzmann hypothesis distribution.

\paragraph{A comprehensive framework for downstream analysis.}
Since the Boltzmann hypothesis distribution assigns counterfactual weights to all observed project profiles, policymakers can compute any mathematically well-defined metric expressible in terms of expectations of functions of the observables, including nonlinear functions of such expectations. Section~\ref{ssc:DownstreamAnalysis} illustrates the flexible scope and substantial practical simplification of this downstream assessment, which contrasts with model-centred approaches. In conventional assessments,  examining a different downstream metric often requires a new model or an extension of an existing one, despite the policy mandate remaining unchanged. Here, the hypothesis counterfactual is already specified, enabling policymakers to shift their effort from deciding which additional models are needed to studying the consequences that matter for their decisions.

Of particular interest in public governance is the \emph{principle of proportionality}, which requires that public intervention does not exceed what is necessary to achieve its objectives. This principle is explicitly recognised in EU law, where Article 5 TEU requires that the content and form of Union action do not exceed what is necessary to achieve the objectives of the Treaties~\cite{TEU_Art5}. In regulatory impact assessment, various policy options are examined in terms of their effectiveness, costs, and consequences~\cite{ECBetterRegulation}. Such assessments presuppose concrete counterfactual scenarios in which the consequences of different policy choices can be examined.

Given specified policy targets, the Boltzmann hypothesis  counterfactual provides such a least-departure scenario for an evidence-based ex-ante assessment of proportionality. By constructing the counterfactual for different policy targets, policymakers can quantitatively compare their implications. If even the minimal overall departure implied by a mandated target entails consequences deemed excessive, policymakers have a clear, quantitative basis for questioning the mandate itself.

\paragraph{Statistical hypothesis testing.}
For a rigorous ex-ante assessment, the intrinsic \(p\)-value allows policymakers to assess the statistical compatibility of the historical baseline with the mandated policy targets under the specified structural conditions. 
Unlike conventional approaches that specify a population model to establish the sampling distribution of a chosen test statistic, the test statistic in our framework emerges from the large-scale behaviour of project repository formation.
The \(\chi^2\) approximation is not imposed independently of the actual process being assessed, but follows from the large-\(N\) regime of effectively independent sampling. The evidentiary meaning of our \(p\)-value is thus grounded in its very sampling-based definition. In our demonstrative application, the intrinsic \(p\)-value quantifies the cumulative probability of sampling a power difference at least as improbable as the historical green-power deficit in the Boltzmann hypothesis  counterfactual.

In the conditional sampling scheme, we retain only alternative repositories whose composition satisfies the specified structural conditions, but is otherwise determined by the combinatorics of a finite sample.  
The uncertainty is thus not about the agreed policy objective or structural conditions that define the hypothesis counterfactual. It arises from fluctuations in the composition of finite-sample repositories around the idealized project composition, which cause the aggregate power difference to depart from the mandated target, while the structural constraints are respected.

Under the Boltzmann hypothesis counterfactual, the intrinsic \(p\)-value aggregates the sampling probability of the historical repository family and all other admissible repository families whose sampling probabilities are no greater than that of the historical family. Consequently, it accounts not only for alternative repositories exhibiting the observed green-power deficit, but also for other power differences belonging to repository families that are at least as improbable as the historical one. (A one-sided test, targeting here green-power deficits, can also be formulated in the Boltzmann framework.)

A \(p\)-value that falls below the acceptance threshold signifies that  the historical green-power deficit would have been highly improbable under the collective influence of real-world mechanisms alone, had we agreed on the green-power surplus that defines the Boltzmann hypothesis  counterfactual. Following the standard logic of hypothesis testing, we reject the hypothesis that such a green-power surplus had been agreed upon over the period covered by the historical data. In other words, the historical baseline can be characterized as statistically incompatible (or implausible) with the mandated green target under the specified structural conditions.

As emphasized by the American Statistical Association~\cite{wasserstein2016statement}, a \(p\)-value does not quantify the probability that a hypothesis is true or false, meaning that scientific conclusions as well as policy decisions should not rest solely on whether it crosses a prescribed threshold. Following the traditional Fisherian logic of hypothesis testing,\footnote{Interestingly, the \(p\)-value, in its very traditional sense, turns out to be the only coarse-grained sampling probability for which we can derive a closed-form expression under the Laplace approximation.} our \(p\)-value likewise does not quantify the probability of obtaining the mandated green-power surplus. Instead, it quantifies the cumulative probability of obtaining the observed deficit, or a more improbable one, under the combined hypothesis of a specific green target. Our contribution is to derive this probability from the collective influence of real-world mechanisms on project repository formation in the large-\(N\) regime, without introducing an ad hoc test statistic or externally specified population model. 

Hence, the result of hypothesis testing in the Boltzmann framework is attributable to the policy objective, alongside the structural conditions, the historical repository, and the lack of more concrete information on project selection. Given this attributability, a significant \(p\)-value (below the acceptance threshold) warrants active restructuring of the historical baseline to achieve the mandated target, should the policymaker decide to implement it following an assessment of downstream consequences.

\section{Results}
\label{sc:Results}

The methodology developed in the preceding section is readily applicable to constructing and assessing counterfactual policy scenarios, given the historical repository and a policy objective. Reweighting observed project profiles subject to the green target and structural conditions (comprising the combined hypothesis of the green mandate) yields the Boltzmann hypothesis counterfactual. From there, the mandate's consequences can be assessed across any dimension already present in the historical repository. Sampling from the Boltzmann hypothesis distribution also provides the $p$-value that quantifies the plausibility of the historical baseline under the mandated policy.

We now apply this statistical framework to the historical electricity-generation projects in the World Bank repository. 
We explore the full range of green-power surpluses that correspond to feasible combined hypotheses. 
Beyond the ex-ante impact assessment of standard fiscal and socio-economic indicators, illustrated through plots, our downstream analysis further assesses the plausibility of the historical green-power deficit under the mandated green-power surpluses.

\subsection{Statistical assessment and fiscal implications of the green transition}
\label{ssc:DesicionTests}

For each targeted power surplus or deficit in Eq.~\eqref{eq:EtaCondition}, we numerically determine the Boltzmann hypothesis distribution $\prob h$ according to
\begin{equation}
	\label{eq:HypothesisDistro}
	\prob h = \argmin_{\prob p\in\mathcal H}\infdiv{\prob p}{\prob f}~,
\end{equation}
where $\mathcal H$ is the hypothesis linear family defined by the combined hypothesis in Section~\ref{ssc:CombinedHypothesis}, meaning that every distribution $\prob p\in\mathcal H$ satisfies Eq.~\eqref{eq:EtaCondition}, Eq.~\eqref{eq:TotalCapacityCondition}, and Eq.~\eqref{eq:HydroCondition}.
In the linear-algebraic language outlined in Section~\ref{ssc:BoltzmannDistro}, this  conditional minimization yields a column vector for each combined hypothesis containing the non-negative weights indexed by the 8-tuples $(\mytuple)$. 

Because the numerical minimization is computationally inexpensive, we evaluate the Boltzmann hypothesis distribution over a dense set of targeted green-brown power differences in Eq.~\eqref{eq:EtaCondition}. This allows the resulting quantities to be effectively displayed as continuous curves without interpolation. The considered range of aggregate power differences includes the historical green-power deficit as well as the entire, feasible green-power surplus regime.

\paragraph{Intrinsic \(p\)-value for the historical green-power deficit.}
Having obtained the Boltzmann hypothesis distribution $\prob h$ for each targeted power difference in Eq.~\eqref{eq:EtaCondition}, we evaluate the intrinsic $p$-value for the observed green-power deficit in Eq.~\eqref{eq:GreenPowerDeficit}. As discussed in Section~\ref{ssc:BoltzmannDistro}, this is approximated by 
the \textsc{cdf} $F_{\chi^2_1}$ of a $\chi^2$ distribution with one degree of freedom, with test statistic
\begin{equation}
	\label{eq:TestStatistic}
	t = 2 N \min_{\prob p\in\mathcal A} \infdiv{\prob p}{\prob h}~.
\end{equation}
Here, $\mathcal A$ denotes the linear family of distributions that support the historically observed green-power deficit, in analogy with the historical repository family.
If $\prob h$ assigns strictly positive weights to all observed project profiles, the test statistic simply reduces to $t=2N\infdiv{\prob f}{\prob h}$. If $\prob h$ assigns zero weight to any observed project profile, the historical repository has zero sampling probability, appearing impossible under the combined hypothesis, rather than merely improbable. 

\begin{figure}[!t]
	\centering
	\includegraphics[width=1.01\linewidth]{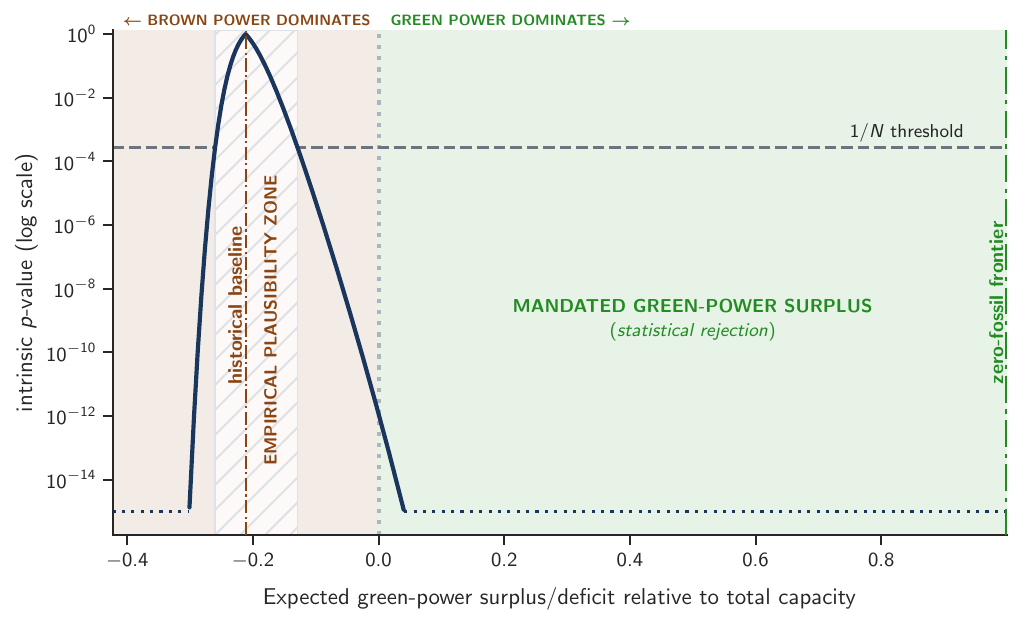}
	\caption{\textbf{Intrinsic $p$-value profile under targeted power differences}. The continuous curve depicts the semi-analytical approximation to the intrinsic $p$-value of obtaining the historical green-power deficit ($-22\,425.6\,\mathrm{MW}$), or a more extreme green-brown power difference, under each hypothesis counterfactual across the considered range of power differences. 
		Computed $p$-values falling below standard machine precision are represented by a dotted horizontal line to avoid displaying exponentially $N$-suppressed fluctuations on the logarithmic scale.}
	\label{fig:pValue}
\end{figure}

Figure~\ref{fig:pValue} shows the resulting $p$-value profile estimated by $1-F_{\chi^2_1}(t)$ across the considered range of mandated power differences, including the historical green-power deficit (for which the $p$-value is trivially one), as well as the complete feasible green-power surplus regime. We identify a plausibility region around the historical baseline consisting of those hypotheses for which the historical green-power deficit is not sufficiently improbable to warrant rejection at the prescribed $1/N$ threshold. This region, thought of as the non-axiomatic analogue of a confidence region, should not be interpreted as indicating which policy was historically implemented or agreed upon. It identifies only the green-power differences that remain statistically compatible with the historical baseline under the specified structural conditions. 

Notice that the plausibility region lies entirely within the range of negative green-power differences. Conversely, the intrinsic $p$-value remains well below the acceptance threshold throughout the entire green-dominance region, where the mandated power difference is positive.\footnote{In our application, the limiting case of vanishing counterfactual probabilities is encountered for stricter green-power surpluses, with individual project profiles already acquiring zero weight before the zero-fossil frontier is reached. At the zero-fossil frontier, all brown project profiles are necessarily phased out. In any event, the $p$-value has already become exponentially suppressed, far below any sensible acceptance threshold, long before reaching such limiting cases.} Following the standard logic of hypothesis testing, we must thus conclude that the historical green-power deficit is statistically incompatible --\,at the $1/N$ threshold\,-- with every tested combined hypothesis involving a green-power surplus. Equivalently, we can reject the hypothesis that a mandated green-power surplus had systematically driven the formation of the historical repository under the specified structural conditions.

According to Section~\ref{ssc:HypothesisTesting}, this statistical assessment has a direct policy interpretation. Under the considered structural conditions, the observed green-power deficit cannot be merely regarded as an ``accidental'' fluctuation arising from administrative inertia and other background processes in an otherwise green-mandated project formation. 
Instead of relying on the continuation of the historical pattern to achieve green dominance, a mandated green target requires deliberate modification of that pattern.

At this point, we clarify the main points that determine the scope of our statistical assessment and its interpretation in the current setting.
\\[-4ex]
\begin{itemize}
	
	\item
	\textbf{Sampling-based formulation:}
	The intrinsic $p$-value is defined as a cumulative sampling probability over repository families. This naturally gives rise to a two-sided formulation, in which repository families are included whenever their coarse-grained sampling probability is no greater than that of the historical repository family, irrespective of whether their power difference lies on the deficit or surplus side. For the present one-dimensional test, a one-sided formulation directed towards green-power deficits could in principle also be constructed without changing our conclusion from hypothesis testing. We nevertheless retain the sampling-based definition because it provides a unified formulation of the intrinsic $p$-value, particularly when multiple mandated conditions are tested simultaneously.
	
	\item
	\textbf{A stringent rejection threshold:}
	We adopt a rejection threshold that scales as $1/N$, rather than the conventional fixed threshold of $5\%$ commonly used in empirical research. Such a fixed threshold is largely a historical convention, mostly attributed to Fisher's proposal of a significance level in the context of relatively small samples. 
	The historical repository of an international organization comprises thousands of projects. When sampling from a hypothesis counterfactual such large alternative repositories, more ``extreme'' deviations can have non-negligible probability. A more stringent threshold, such as $1/N$, takes this feature of sampling-based $p$-values into account. 
	Importantly, the conclusions reported above do not depend on this stringent choice, since the computed $p$-values throughout the green-dominance region remain far below any conventional, much higher threshold.
	
	\item
	\textbf{An integrated historical horizon:}
	Our analysis integrates all historical records containing valid, green-classifiable projects over the forty-year period from 1984 to 2024 into a single historical repository. This extended horizon reduces the influence of fluctuations that may arise within individual years and increases the sample size available for the  statistical assessment. It also provides an increased number of observed project profiles. In particular, the broader range of green-to-brown power differences contained in the historical repository makes a wider range of combined hypotheses feasible within the observed project profiles than would be possible with substantially shorter historical periods. In total, our statistical assessment  concerns the aggregate formation of the historical repository over the entire forty-year horizon, rather than fluctuations in any particular year.
	
	\item
	\textbf{Structural conditions:}
	The hypothesis testing is performed under the structural conditions of fixed total nameplate capacity and fixed hydroelectric capacity, both calibrated to their historical levels, as discussed in Section~\ref{ssc:CombinedHypothesis}. These conditions impose physical and operational constraints on the counterfactual scenario. In particular, changing total nameplate capacity would change the scale of the historical commitment, while hydroelectric capacity is subject to technological, operational, physical, and environmental constraints on the availability of suitable water resources.
	Although our statistical rejection of green-power surplus hypotheses is subject to these structural conditions, the Boltzmann framework allows such conditions to be lifted, replaced, or extended when the policy question or available information warrants doing so.
\end{itemize}

\paragraph{Shifts in investment allocation.}
Having established that the historical green-power deficit is statistically incompatible with any mandated green-power surplus, we now turn to a green-mandate's fiscal implications. Following this significant result of the statistical assessment, the relevant question is no longer whether a green transition could naturally emerge from administrative inertia and the broader real-world processes shaping the historical repository. As such an emergence is exceedingly improbable, active intervention is required to change the historical pattern. The ex-ante policy exploration now concerns the reallocation of investment that would accompany the green transition under the specified structural conditions and governance principle.

To investigate fiscal consequences, we examine the $X$ observable describing a project's total investment, as outlined in Section~\ref{ssc:Observables}. Notice that this observable has not been part of the combined hypothesis of the green mandate that we formulated in Section~\ref{ssc:CombinedHypothesis}. However, its analysis is directly relevant to the principles and framework of {green budgeting}, where budgetary allocation is used as a policy instrument to advance environmental objectives.
In the hypothesis counterfactuals constructed for each mandated power difference, the aggregate effect of investment reallocation due to the green transition is captured by moments of $X$.  

For each targeted green-brown power difference, we track the share of the total expected investment that is allocated to each technology tag in the Boltzmann hypothesis counterfactual. For a given technology class $s$, this share is given by
\begin{equation}
	\label{eq:TaggedBudget}
	\sum_x x\,h(x,s), \qquad s\in\left\{\texttt{green,brown}\right\},
\end{equation}
relative to the overall expected project cost,
\begin{equation}
	\label{eq:TotalBudget}
	\sum_x x\,h(x) ~.
\end{equation}
The  weights in these expressions are those of the Boltzmann hypothesis distribution obtained from the information-divergence minimization in Eq.~\eqref{eq:HypothesisDistro}. By construction, the green and brown investment shares sum to unity ($100\%$) for every targeted power difference considered.

\begin{figure}[!t]
	\centering
	\includegraphics[width=1.\linewidth]{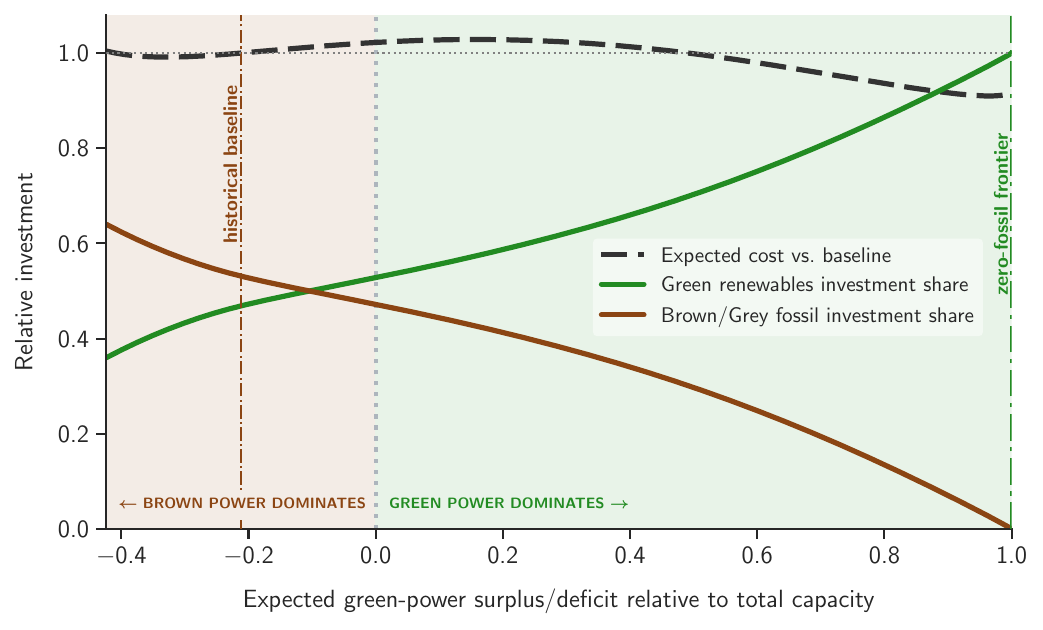}
	\caption{\textbf{Investment allocation and project cost under targeted power differences}. The continuous curves depict the expected investment shares allocated to the green and brown technology tags. The investment shares cross in the brown-dominance region, prior to power parity (zero power difference).
	The dashed curve depicts the expected project cost (normalized by its historical baseline), which varies non-monotonically across the targeted power differences.}
	\label{fig:Budget}
\end{figure}

The trajectories in Figure~\ref{fig:Budget} reveal an asymmetry between tagged investment allocation and tagged nameplate capacity. At the historical baseline, the majority of investment is directed toward brown projects. At the investment crossover, where spending on green and brown projects is equal, brown projects nevertheless still account for a larger share of total capacity (negative power difference). Investment parity thus occurs prior to power parity along the transition toward green dominance. Within the Boltzmann hypothesis counterfactual, a green-power surplus consequently requires an allocation of slightly more than half of the expected investment to green projects. 

Interestingly, the expected cost per project in Eq.~\eqref{eq:TotalBudget} follows a non-monotonic trajectory across the mandated power differences. To facilitate comparison with the historical baseline, we have normalized this quantity by its historical value, obtained for $\prob h=\prob f$ in Eq.~\eqref{eq:TotalBudget}. As we move from the historical baseline into the green region, the expected project cost initially increases, to roughly $3\%$ above the historical baseline. Deeper in the green-power dominance region, however, this trend reverses, and the expected cost falls substantially, to approximately $9\%$ below the historical baseline near the zero-fossil frontier. Thus, increasingly stringent green-power mandates do not correspond to a monotonically increasing expected project cost within the counterfactual construction.

Taken together, these results provide two considerations directly relevant to green budgeting. First, \textit{parity in investment allocation does not imply parity in tagged power}. Under an otherwise minimal transition from the historical baseline, equal allocation of investment between green and brown projects is reached before balancing of their power contributions. Second, from a purely fiscal perspective, a weaker green-power surplus appears undesirable, while pursuing a stronger green-power surplus seems more beneficial, as it allows the green transition to lower the expected project cost even below the historical baseline. In other words, the expected project cost trajectory alone suggests that policymakers could push more aggressively for the green transition.

Plots~\ref{fig:pValue} and~\ref{fig:Budget} naturally share the same formatting pattern.
The statistical assessment and the fiscal analysis rely on the same counterfactual construction, with the former assessing the plausibility of the historical baseline under the mandated power difference and the latter quantifying the changes in investment allocation associated with the minimal mandated transition. At the technical level, the same Boltzmann hypothesis distribution $\prob h$, evaluated for each power difference in Eq.~\eqref{eq:EtaCondition}, enters both the test statistic in Eq.~\eqref{eq:TestStatistic} and the budgetary metrics in Eqs.~\eqref{eq:TaggedBudget} and~\eqref{eq:TotalBudget}.

\subsection{Further implications of the green transition}
\label{ssc:DownstreamAnalysis}

We now turn to dimensions of the hypothesis counterfactuals that concern project composition, technological allocation, and development-related disparities. Additional analyses are presented in Appendix~\ref{app:DownstreamAnalysis}.

\paragraph{Project prevalence across the green transition.}
\begin{figure}[!t]
	\centering
	\includegraphics[width=1.\linewidth]{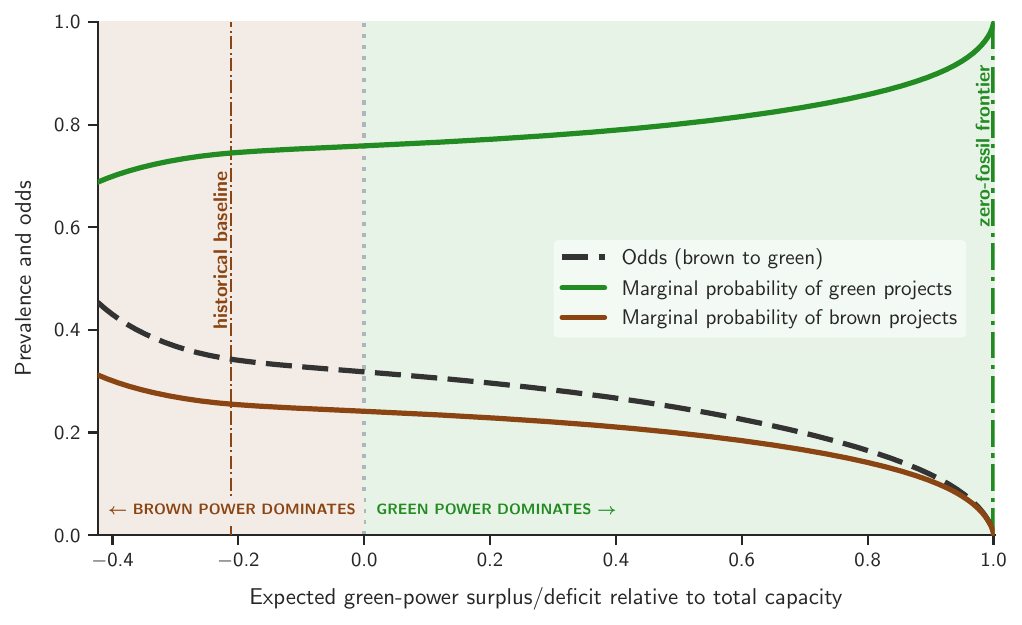}
	\caption{\textbf{Project prevalence across targeted power differences}. The curves depict the expected relative frequencies (counterfactual probabilities) of the green and brown technology tags across the targeted power differences, together with the corresponding selection odds.}
	\label{fig:Prevalence}
\end{figure}
 We first consider the prevalence of green- and brown-tagged projects across the targeted power differences. In Figure~\ref{fig:Prevalence}, the trajectories of $h(s)$ for $s\in\{\texttt{green,\,\,brown/grey\,fossil}\}$ show how the tagged composition of the hypothesis counterfactual changes as the mandated power difference is varied.
Remarkably, moving from the historical green-power deficit to exact green-brown power balance requires only a $1.4\%$ increase in the relative frequency of green projects. The transition to power balance can therefore initially be achieved with only a modest change in the  prevalence of the tagged technology classes. 

The relative stability of the prevalence trajectories reflects the regularizing effect of the minimal-departure counterfactual. Rather than preferentially assigning zero probability to individual project profiles to achieve the mandated power difference, the $I$-projection redistributes probability mass across the observed profiles, preserving the historical composition as far as the imposed conditions permit. Only when the mandate itself requires zero brown power, at the zero-fossil frontier, are brown projects eliminated altogether.

\paragraph{Granular resource allocation and capital efficiency.}
Green tagging is an important prerequisite for green budgeting, but it does not reveal how expenditure is allocated across green technologies or what economic characteristics accompany that allocation. An ex-ante assessment should go beyond the volume of expenditure classified as green to examine such aspects.

Our counterfactual construction provides this additional layer of analysis without directly requiring any technology-specific policy preferences. Because the Boltzmann hypothesis distribution assigns counterfactual probabilities to the complete project profiles, any quantity computable from the historical repository can likewise be evaluated in the corresponding hypothesis counterfactual. 
Specifically, we examine the fiscal and economic consequences using the $Y$ and $X$ observables in a downstream analysis stratified by the granular variable $\tilde S$ introduced in Eq.~\eqref{eq:GreenTechnologies}.

\begin{figure}[!t]
	\centering
	\includegraphics[width=1.0\linewidth]{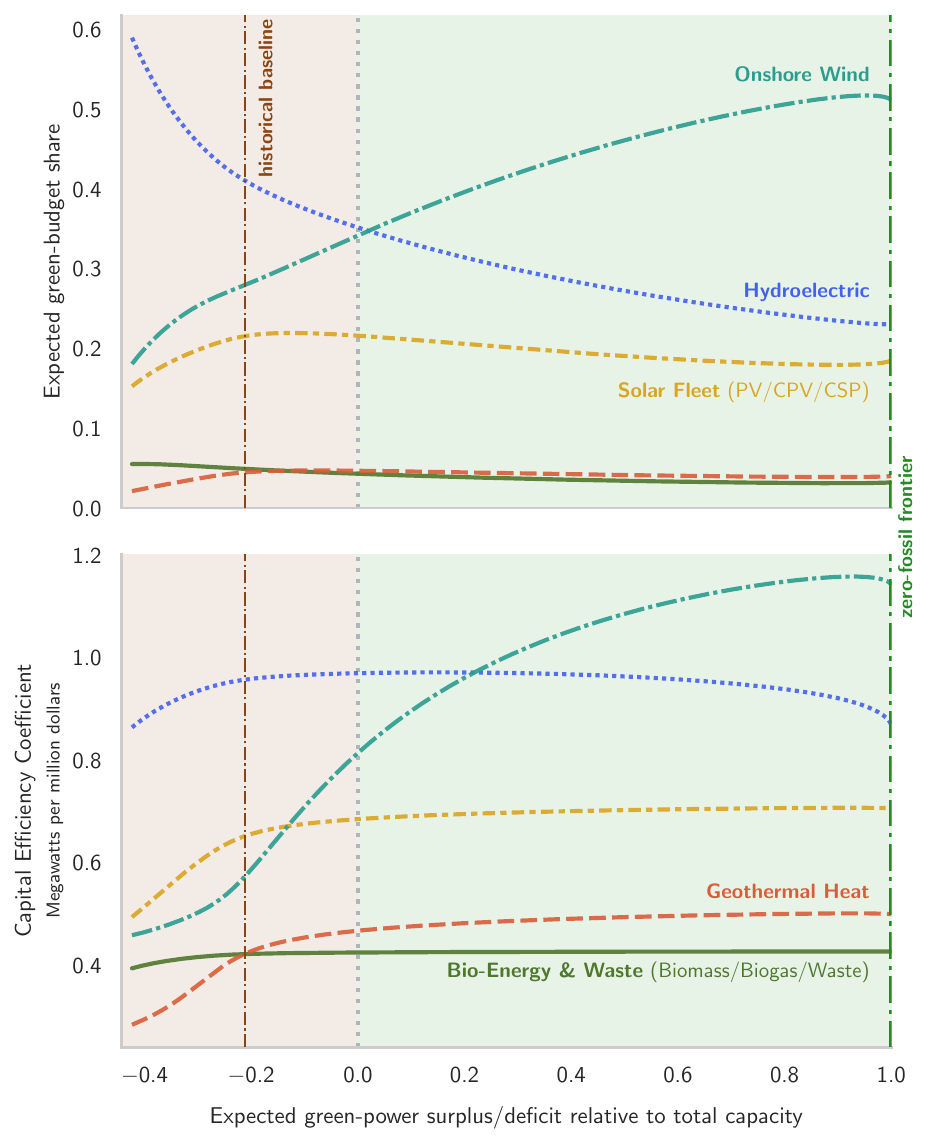}
	\caption{\textbf{Green investment and capital efficiency across targeted power differences}. The upper panel depicts the green-investment shares of the five green-tagged technologies, showing the transition in investment dominance from hydroelectric power to onshore wind. The lower panel displays the corresponding capital efficiency coefficients, showing the sharp rise of onshore wind against the comparatively stable trajectories of the remaining technologies. The same color scheme is used between the two panels.}
	\label{fig:GreenBreakdown}
\end{figure}

These consequences are quantified by granular aggregate investment (first moment of $X$),
\begin{equation}
	\label{eq:GranularMean_X}
	\sum_x x\,p(x, s=\texttt{green}, \tilde s)~,
\end{equation}
which we normalize by the green aggregate investment
\begin{equation}
	\sum_{x} x\,p(x, s=\texttt{green})= \sum_{\tilde s\neq \texttt{N/A}}\sum_{x} x\,p(x, s=\texttt{green}, \tilde s)
\end{equation}
to obtain the normalized green-investment shares shown in the upper plot of Figure~\ref{fig:GreenBreakdown}. This ensures that the five curves in the upper plot sum to unity at every targeted power difference.
As an economic metric, we use the capital efficiency coefficient shown in the lower plot of Figure~\ref{fig:GreenBreakdown}, defined as the nameplate capacity of a green-technology class per unit of its aggregate investment. In the Boltzmann-hypothesis counterfactual, it is given by the ratio of granular aggregate capacity (first moment of $Y$),
\begin{equation}
	\label{eq:CapitalEfficiencyCoefficient}
	\sum_y y\, p(y, s=\texttt{green}, \tilde s) 
	~,
\end{equation}
to the granular aggregate investment in Eq.~\eqref{eq:GranularMean_X}.

In Figure~\ref{fig:GreenBreakdown}, the trajectories reveal how the green investment allocation changes across the mandated power differences. At the historical baseline, hydroelectric projects account for the largest share of aggregate green investment, while the transition toward green-power dominance progressively shifts the allocation toward onshore wind. The remaining green technologies account for smaller --\,marginally fluctuating\,-- shares throughout the transition. In particular, solar fleet receives a substantial green-investment share across the transition, while geothermal heat and bio-energy \& waste retain minimal funding shares.

The capital-efficiency trajectories provide a complementary economic perspective to the fiscal reallocation. The increasing green-investment share of onshore wind is accompanied by a pronounced increase in its capital-efficiency coefficient, whereas the corresponding coefficient for hydroelectric projects remains comparatively stable. The other technologies exhibit lower capital-efficiency coefficients with less pronounced changes across the mandated green transition.

The fixed hydroelectric-capacity condition in Eq.~\eqref{eq:HydroCondition} is important for interpreting these trajectories. If left unconstrained, hydroelectric power would dominate both green investment and the capital-efficiency coefficient in the Boltzmann hypothesis counterfactual under a stringent green-power surplus. Such dominance is not a methodological artifact. Instead, it is a mere consequence of the scale of hydroelectric power generation, with dams constituting large infrastructure projects that provide exceptionally large nameplate capacities. As explained below Eq.~\eqref{eq:HydroCondition}, the relevant exogenous constraint for our ex-ante analysis is that this historical scale of hydroelectric power cannot be extrapolated indefinitely into the future. Hence, we chose to keep aggregate hydroelectric capacity fixed across the green transition.\footnote{From a purely methodological perspective, the Boltzmann framework allows one to fix the hydroelectric aggregate to any value --\,supported by observed nameplate capacities\,-- away from the historical baseline.}

All in all, the granular fiscal and economic trajectories emerge from minimal adaptation to the combined hypothesis conditions, which involve neither the fiscal observable $X$ nor the granular observable $\tilde S$. In this way, the Boltzmann hypothesis counterfactual pinpoints onshore wind, followed by hydroelectric and then solar fleets, as the driver technologies of an increasingly stringent green transition under the designated exogenous constraints.

\paragraph{International disparities.}
To examine the development-related dimension of investment efficiency, we analyze the capital efficiency coefficient stratified by the secondary observable $Z_3$, corresponding to the World Bank's International Development Association (\textsc{ida}) eligibility status. For each development category $z_3$, the coefficient (cf.\ Eq.~\eqref{eq:CapitalEfficiencyCoefficient} and Eq.~\eqref{eq:GranularMean_X}) is given by the ratio of aggregate nameplate capacity to aggregate investment:
\begin{equation}
	\label{eq:CapitalEfficiency_bhyIDA}
	\sum_y y\, h(y\vert z_3) \bigg/ \sum_x x\, h(x\vert z_3) ~.
\end{equation}
Recall that $\prob h$ has already been evaluated at each targeted power difference.

\begin{figure}[!t]
	\centering
	\includegraphics[width=1.\linewidth]{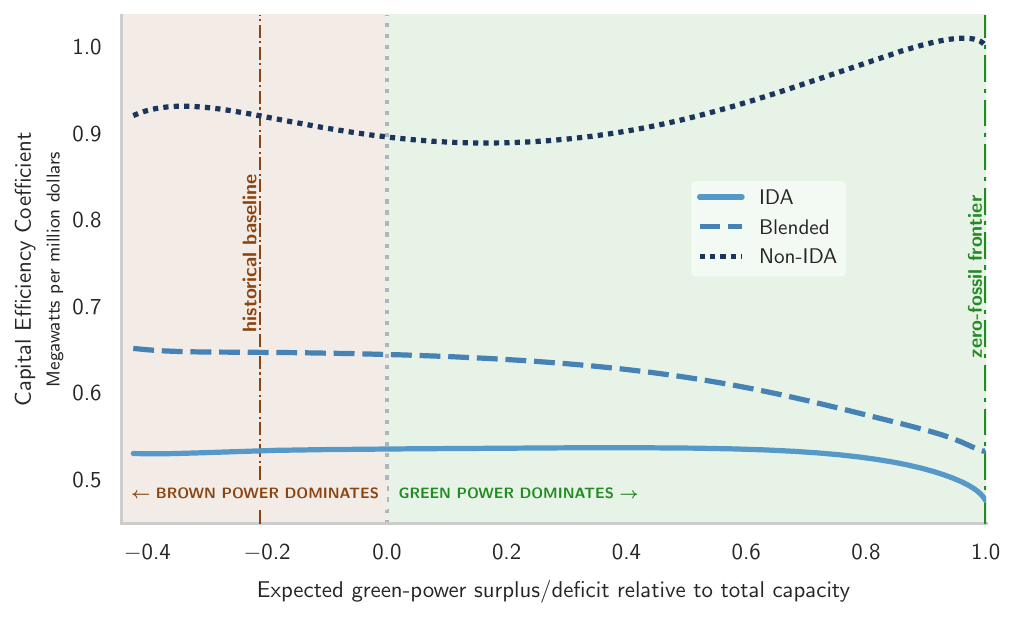}
	\caption{\textbf{Capital efficiency by {IDA} status}. Depicted is the ratio of aggregate nameplate capacity to aggregate investment, defined in Eq.~\eqref{eq:CapitalEfficiency_bhyIDA}, stratified by \(Z_3\) across the targeted power differences.}
	\label{fig:CapitalEfficiany_IDAstatus}
\end{figure}

Figure~\ref{fig:CapitalEfficiany_IDAstatus} shows a persistent difference in capital efficiency across the three development categories. Non-\textsc{ida} countries maintain the highest capital efficiency coefficient, which fluctuates across the targeted power differences and reaches a local minimum within the green-dominance region, while remaining above $0.88\,\CECunit$. By contrast, low-income \textsc{ida}-eligible and blend countries follow either nearly stagnant or systematically declining trajectories, with both groups falling below $0.6\,\CECunit$ as the counterfactual approaches the zero-fossil frontier.

The stratified results have important implications for the interpretation of the observed capital-efficiency gap. The \textsc{ida} status distinguishes countries according to their financing and development conditions: \textsc{ida}-eligible countries are classified by the World Bank on the basis of low per-capita income and limited creditworthiness for borrowing on market terms, while blend countries combine \textsc{ida} eligibility with some access to \textsc{ibrd} financing; non-\textsc{ida} countries are eligible only for \textsc{ibrd} borrowing within this classification~\cite{WorldBankPPI_CountryClassification}. Assuming that these distinctions serve as a proxy for, among other things, differences in the underlying infrastructure and investment base, the historical differences in capital efficiency observed across the three groups indicate a pronounced distributional disparity in the economic capacity to translate invested capital into nameplate capacity.

Since we have not directly controlled for such disparities, the Boltzmann-hypothesis counterfactual propagates the historical capital-efficiency gap throughout the mandated transition. More importantly, the aggregate gap between the three groups becomes increasingly pronounced as the repository approaches the zero-fossil frontier. This behavior is particularly concerning from a policy perspective. If increasingly stringent green-power surpluses are imposed without additional structural conditions, the transition can amplify rather than reduce an existing disparity in the efficiency with which invested capital translates into nameplate capacity.

The closing part of our downstream analysis underscores that a green-power mandate on its own does not address the observed disparity in capital efficiency among countries. 
To prevent the green transition from reinforcing existing disparities, the policy objective must be supplemented with additional structural conditions aimed at addressing the historical disparity in capital efficiency. 
Fortunately, the Boltzmann framework allows for a seamless extension of the combined hypothesis to incorporate such equity-oriented conditions.


\section{Conclusions and outlook}

\subsection{Conclusions}
\label{sc:Conclusions}

Taking an orthogonal approach to conventional statistics, we provide a tractable solution to the ex-ante impact assessment of policy objectives by explicitly constructing the Boltzmann hypothesis counterfactual. The composition of this counterfactual scenario is solely attributable to the historical project repository, policy objectives, and any relevant constraints directly meaningful to the policymaker.

Specifically, we consider policy objectives whose combined hypothesis can be expressed through aggregate conditions constraining the corresponding generalized moments. We focus on settings where nothing concrete is known or otherwise assumed about the actual process governing project selection and repository formation. In our previous work, we established through large-\(N\) saddle-point techniques that the physics of such a maximally agnostic sampling process supplies the ingredients that Kolmogorov's framework seeks to derive axiomatically. In particular, coarse-graining the sampling probability inevitably yields an approximately tractable, intrinsic \(p\)-value in the most traditional Fisherian sense.

Retaining the statistical physics perspective, we show that a 
historical repository comprising variables of arbitrary modality, together with the combined hypothesis of the policy objective, suffices to uniquely construct the Boltzmann hypothesis counterfactual. Restricting the analysis to observed project profiles only,  this uniqueness follows from epistemic consistency of the intrinsic $p$-value. 
The Boltzmann hypothesis counterfactual provides a statistical assessment of the policy objective through the intrinsic \(p\)-value, while also evaluating any downstream fiscal or socioeconomic metric computable in the original historical repository. Remarkably, the least-departure property of this  hypothesis counterfactual that is selected by epistemic consistency offers a solid basis for assessing proportionality.

Whenever the large-\(N\) regime applies, any self-consistent statistical or downstream ex-ante assessment of a policy objective that claims to be attributable solely to the historical baseline and the combined hypothesis must agree with our leading-order results, irrespective of the statistical framework used. By construction of the Boltzmann hypothesis counterfactual, any leading-order disagreement within this regime necessarily reveals that the alternative assessment, consciously or otherwise,
“corrects” the historical repository composition by incorporating assumptions about project selection, further administrative aspects etc. Such additional assumptions are not inherently objectionable, provided they are explicitly acknowledged. 

As a proof of principle, we have used publicly available World Bank data to construct the Boltzmann hypothesis counterfactual across the feasible range of mandated green-brown power differences. The exceedingly low intrinsic $p$-value for any feasible green-power surplus means that the historical green-power deficit would have been highly improbable solely due to non-systematic real-world influences\footnote{In the large-scale description, non-systematic real-world influences are effectively captured by the sampling uncertainty in Boltzmann's illustrative Gedankenexperiment. This uncertainty is a physical counterpart to randomness from Kolmogorov's formulation.} on repository formation, if the green mandate had been in effect.
According to the conventional lore of statistical hypothesis testing, we thus conclude that no green mandate had been in effect, so that active intervention to modify historical patterns is needed to achieve any level of green-power surplus deemed desirable. 

The downstream analysis reveals that parity in green and brown investment precedes parity in green and brown power, while the total average project cost follows a non-monotonic trajectory across the mandated power differences. Private ownership and project vintage remain comparatively stable. At the technology level, the Boltzmann hypothesis counterfactual identifies onshore wind as the eventual driver of the green transition. Regional investment allocations remain largely stable, with mandated shifts absorbed primarily through within-region reallocations of funding from brown to green power. Differences in capital efficiency across \textsc{ida}, blend, and non-\textsc{ida} economies persist and widen as the green mandate becomes more stringent.

The World Bank application illustrates three practical consequences of our framework. First, the intrinsic $p$-value assesses the statistical compatibility of the historical baseline with the mandated green transition, while the downstream fiscal and socioeconomic consequences inform the subsequent policy decision. Second, the same Boltzmann hypothesis counterfactual evaluates these downstream consequences for each feasible combined hypothesis, without requiring new statistical or econometric models. Third, the efficient computation of the Boltzmann hypothesis distribution and the closed-form approximation of the intrinsic $p$-value allow thousands of mandated targets to be assessed within seconds. 

In this way, the proposed framework provides comprehensive quantitative support for ex-ante green budgeting. This is consistent with the European Union Green Budgeting Reference Framework's emphasis on policy assessment, transparency, and accountability~\cite{EUGreenBudgetingReferenceFramework}. The implications of meeting a mandated target across fiscal, economic, technological, and geographical dimensions can inform the evaluation of policy compatibility with applicable fiscal and economic constraints. In addition, such quantitative assessments can contribute to green medium-term fiscal frameworks, which seek to incorporate the economic and fiscal consequences of climate policies into strategic public-finance planning~\cite{Caselli2024}.

Inspired by statistical physics, the payoff of our large-scale reasoning is that we can rapidly assess within the same construction both the downstream consequences and the statistical (im)probability of the historical deficit under the combined hypothesis of a green mandate. Without ever relying on a population or microscopic physical model of the processes governing project-repository formation, this self-consistent assessment is readily attributable to the historical repository and the combined hypothesis that the policymaker formulates.

\subsection{Outlook}
\label{sc:Outlook}

Although demonstrated on a dataset extracted from the World Bank repository, the Boltzmann construction facilitates wider application. It can equally be applied to project repositories maintained by national ministries or other public authorities, enabling assessments at the level of individual countries or groups of countries, such as EU Member States. Counterfactual weights are inexpensive to compute using a standard optimized Newton-Raphson algorithm, while the intrinsic $p$-value can be approximated using the corresponding $\chi^2$ statistic, without rejection sampling. The Boltzmann hypothesis counterfactual can therefore be efficiently recomputed as projects are added, technology tags reclassified, or the green mandate updated.

When the historical repository includes a temporal observable, successive historical windows provide a retrospective direction of analysis. A Boltzmann hypothesis counterfactual constructed from an earlier window given a mandate can be compared with downstream outcomes observed in a later window realizing this mandate. Discrepancies can then help identify which additional policy targets or structural conditions would have been required for the earlier counterfactual to reproduce the later outcomes, providing a means of tracing actual policy interventions or further systematic effects over the observed period.

The construction is fundamentally domain-agnostic; in particular, the modality of the original attributes becomes immaterial in the effective description of generic sampling. This makes it applicable wherever low-dimensional societal objectives must be reconciled, under minimal restructuring, with high-dimensional portfolios of assets, programs, or other units. Potential applications include resource allocation in international humanitarian responses, including migrant and refugee crises, and demographic representation in selection procedures, such as persistent gender imbalances in national and regional labor structures~\cite{loukas2023demographicparitymitigatingbiases}. The linear-algebraic formulation also accommodates multiple policy objectives, for instance by incorporating fairness conditions alongside a green target.

A methodological extension concerns the form of the policy objective. As a first insight from this orthogonal approach, we addressed combined hypotheses that constrain expectations of functions of the observables to specified values, yielding linear conditions on generalized moments. Extending the construction to nonlinear conditions on expectations would accommodate more complex algebraic relations among generalized moments and thus more complex policy questions.

A further extension concerns repository formation. In a deliberately agnostic spirit, the present construction treats this process effectively as multinomial, with unknown real-world influences absorbed into generic-sampling uncertainty. Where systematic features of the actual project selection are known and independently justified, they could instead be incorporated either through generalized-moment conditions or through generalized sampling schemes beyond multinomial sampling~\cite{NivenGrendar2009}. This would allow additional information about repository formation to enter the counterfactual construction in a controlled manner without requiring a fully specified model of the underlying socioeconomic, technological, or administrative mechanisms.

\vspace{0.6cm}

\subsection*{Data Availability}
\vspace{-0.2cm}
The data used in this study are publicly available from the World Bank Private Participation in Infrastructure (\textsc{ppi}) database. The accompanying Jupyter notebook \href{https://gitlab.uni-marburg.de/fb20/ag-chung/greenboltzmann}{https://gitlab.uni-marburg.de/fb20/ag-chung/greenboltzmann} provides the complete computational workflow from data processing and designation of the historical repository to the construction of Boltzmann hypothesis counterfactuals and subsequent analysis. This includes the computation of intrinsic $p$-values and downstream metrics used to generate the figures in the paper.

\subsection*{Acknowledgments}
\vspace{-0.2cm}
The work of K.S.\ and G.F.L.\ is supported by the Jean Monnet Chair ``Economic and Fiscal Governance Framework and Transparency in EU'' (EUFiscalGov), funded by the European Union under the Erasmus+ programme (Project No.~101239868).
\vspace{-0.4cm}
\begin{figure}[!h]
    \hfill
    \includegraphics[width=0.31\linewidth,trim=0 0 62 0,clip]{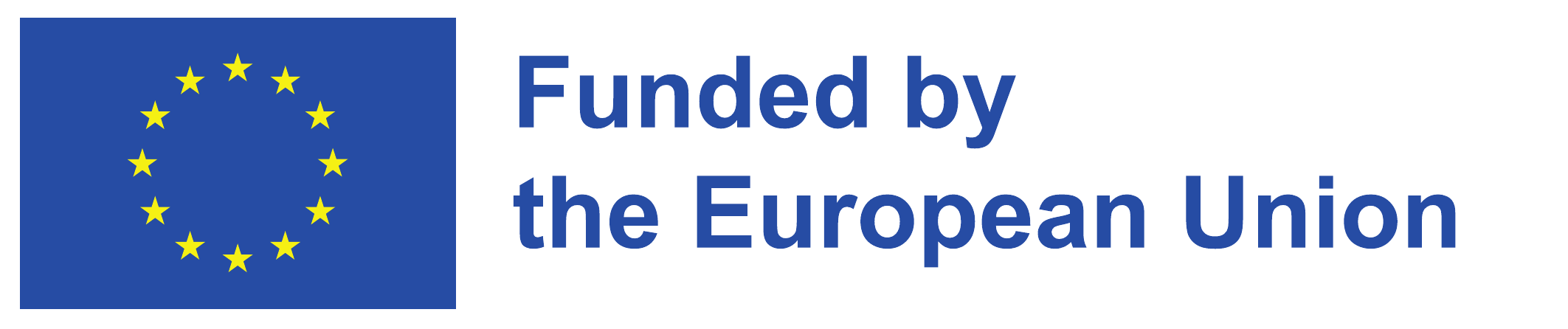}
\end{figure}
\vspace{-1.3cm}
\subsection*{Disclaimer}
\vspace{-0.2cm}
The views and opinions expressed are solely those of the authors and do not necessarily reflect those of the European Union or the European Education and Culture Executive Agency. Neither the European Union nor the granting authority can be held responsible for them.

\printbibliography
\vspace{0.5cm}
\sectionlinetwo{purple}{88} 
\vspace{0.5cm}

\appendix
\section{Appendix}


\subsection{Summary of the Boltzmann statistical framework}
\label{app:BoltzmannFrameworkSummary}
\vspace{-0.3cm}

\renewcommand{\arraystretch}{1.05}
\setlength{\LTcapwidth}{\textwidth}

\begin{longtable}{p{0.29\textwidth} !{\color{gray!50}\vrule width 0.5pt}  p{0.71\textwidth}}
	
	\caption{\textbf{Interdisciplinary terminology from statistical physics in public governance}. Statistical structure emerges by coarse-graining the selection probability of alternative project repositories within the universality class of effectively independent sampling. For a schematic depiction, see Figure~\ref{fig:EffectiveProjectSelection}.}\\
	\textbf{Concept / construction} & \textbf{Description / explanation} \\
	\hline 
	\endfirsthead
	
	\multicolumn{2}{l}{\small{Summary table 
		 --- continued from previous page}}\\
	\hline
	\textbf{Concept / construction} & \textbf{Description / explanation} \\
	\hline
	\endhead
	
	\hline
	\multicolumn{2}{r}{\small{Continued on next page}}\\
	\endfoot
	
	\hline
	\endlastfoot
	
	\entryspace Sample size $N$
	&
	The number of historically recorded sampling units, which defines the scale of finite-sample uncertainty and hence the applicability of the Boltzmann statistical-physics perspective.
	\\[1ex]
	
	\entryspace Infrastructure projects 
	&
	They constitute the sampling units, the abstract balls in the Boltzmann urn. These are the objects that are repeatedly drawn in the Gedankenexperiment and whose observed characteristics determine the project profiles.
	\\[0.8em]
	
	\entryspace Observables
	&
	Selected project attributes relevant to the ex-ante analysis, such as invested budget, private investment share, nameplate capacity, technology, or geographical characteristics. An observable is not declared a priori to be metric or categorical; its modality follows from the form of conditions imposed on it.
	\\[0.8em]
	
	\entryspace Project profiles
	&
	A project profile is the joint specification of the selected observables for an individual project. Profiles encode the possible combinations of project characteristics that distinguish the types of balls in the Boltzmann urn.
	\\[0.8em]
	
	\entryspace Observed project profiles
	&
	Only valid project profiles recorded in the available data are admitted, preventing the counterfactual construction from extrapolating to projects with unreasonable characteristics. The finite set of observed profiles is bounded by $N$. 
	\\[0.8em]
	
	\entryspace Historical repository
	&
	The finite collection of the $N$ historically observed projects, represented by their distinct project profiles and corresponding counts.
	\\[0.8em]
	
	\entryspace Empirical \newline distribution $\prob f$
	&
	Obtained directly from the historical repository, it is the distribution whose probabilities are given by the relative frequencies of observed project profiles. 
	\\[0.8em]
	
	\entryspace Counterfactual \newline distribution $\prob p$
	&
	A distribution assigning non-negative weights based on some counterfactual scenario to the observed project profiles. 
	\\[0.8em]
	
	\entryspace Counterfactual repository
	&
	An infinitely large repository constructable from the observed project profiles. Its composition is dictated by the counterfactual distribution $\prob p$. 
	\\[0.8em]
	
	\entryspace Generic sampling
	&
	Effective description of project selection in the large-sample regime. It expresses non-systematic fluctuations in the prevalence of project profiles due to many unknown influences through sampling uncertainty. Conceptualized by Boltzmann's Gedankenexperiment, where a collection of \(N\) projects is  independently drawn from a counterfactual repository.
	\\[0.8em]
	
	\entryspace Structural conditions
	&
	Aggregate conditions imposing prescribed values on generalized moments, i.e.\ expectations of functions of observables, independently of the particular green target. In the present application, these include fixed total and hydroelectric capacities, together with the normalization associated with the fixed repository size.
	\\[0.8em]
	
	\entryspace Green-mandate condition
	&
	Mean difference condition specifying the mandated green-power surplus.	
	Depending on the policy objective, multiple mandated conditions can be simultaneously imposed.
	\\[0.8em]
	
	\entryspace Combined hypothesis
	&
	The set of structural and green-mandate conditions. It expresses through the concrete values of aggregates all directly controllable dimensions in the policy question. 
	\\[0.8em]
	
	\entryspace Hypothesis distribution 
	&
	A counterfactual distribution that satisfies a feasible combined hypothesis
	\\[0.8em]
	
	\entryspace Hypothesis counterfactual
	&
	A counterfactual repository whose composition is described by a hypothesis distribution. 
	\\[0.8em]
	
	\entryspace Alternative repositories
	&
	Project repositories of the same size $N$ as the historical repository, sampled from a hypothesis counterfactual under the structural conditions. They represent alternative historical versions in the policy universe of the hypothesis counterfactual.
	\\[0.8em]
	
	\entryspace Repository families
	&
	Alternative repositories are grouped according to their aggregate green-to-brown power difference. Distinct alternative repositories can therefore belong to the same family when they exhibit the same aggregate power difference alongside the same structural generalized moments.
	\\[0.8em]
	
	\entryspace Historical \newline repository  family
	&
	The repository family defined by the historically observed green-power deficit. The historical repository itself is one member of this family and, by construction of the Boltzmann hypothesis distribution, is its most probable member.
	\\[0.8em]

	\entryspace Boltzmann hypothesis\newline distribution $\prob h$
	&
	The $I$-projection of $\prob f$ onto the hypothesis linear family.   
	Attributability of the counterfactual scenario to the empirical structure not directly specified by the combined hypothesis unambiguously fixes the hypothesis distribution to this projection.
	\\[0.8em]
	
	\entryspace Boltzmann\newline hypothesis counterfactual
	&
	The hypothesis counterfactual whose composition is dictated  by the Boltzmann hypothesis distribution, realizing the policymaker's counterfactual scenario with the least departure from the historical repository. 
	Under this counterfactual, the historical repository is the most probable within  the historical repository family.
	\\[0.8em]
	
	\entryspace Intrinsic $p$-value
	&
	The cumulative probability over the historical repository family and all repository families that are no more probable under generic sampling from the Boltzmann hypothesis counterfactual. It quantifies how plausibly the historically observed green-power deficit, or a more extreme power difference, can arise under a specific mandated green-power surplus.
	\\[0.8em]
	\hline
\end{longtable}

\begin{figure}[h]
	\centering	
	\begin{tikzpicture}[
		>=Stealth,
		input/.style={
			rectangle,
			rounded corners=2pt,
			draw=black!65,
			fill=gray!8,
			text width=4.4cm,
			align=center,
			minimum height=1.35cm,
			inner sep=6pt
		},
		central/.style={
			rectangle,
			rounded corners=3pt,
			draw=black,
			fill=gray!18,
			text width=3.5cm,
			align=center,
			minimum height=1.5cm,
			inner sep=7pt,
			font=\bfseries
		},
		output/.style={
			rectangle,
			rounded corners=2pt,
			draw=black!65,
			fill=gray!8,
			text width=3.5cm,
			align=center,
			minimum height=1.35cm,
			inner sep=6pt
		},
		flow/.style={->, thick, draw=black!75}
		]
		
		\node[input] (repository) at (0,0) {
			Historical\\repository of \(N\) projects
		};
		
		\node[input, below=0.35cm of repository] (conditions) {
			\(d\) conditions on
			generalized moments of observables
		};
		
		\node[input, below=0.35cm of conditions] (knowledge) {
			Lack of knowledge\\ about
			project-selection processes
		};
		
		\node[
		draw=black!45,
		dashed,
		rounded corners=4pt,
		fit=(repository)(conditions)(knowledge),
		inner sep=0.3cm,
		label={[font=\bfseries]above:Inputs}
		] (inputs) {};
		
		\node[central,text=bostonuniversityred] (counterfactual) at (6,-2) {
			Boltzmann hypothesis\\counterfactual
		};
		
		\node[output] (assessment) at (10.9,-1) {
			Statistical\\assessment
		};
		
		\node[output] (implications) at (10.9,-3) {
			Downstream policy\\implications
		};
		
		
		\draw[flow]
		(inputs.east |- counterfactual.west)
		--
		(counterfactual.west)
		node[midway, above, font=\small] {\(d\ll N\)};
		
		
		\draw[flow] (counterfactual.east) -- ++(0.6,0)
		|- (assessment.west);
		
		\draw[flow] (counterfactual.east) -- ++(0.6,0)
		|- (implications.west);
		
	\end{tikzpicture}
	\caption{\textbf{Flow chart for the Boltzmann framework.} From the minimal input of a policy question expressible through generalized moments of observables in the historical repository to a comprehensive ex-ante impact assessment.}\label{fig:FlowChart}
\end{figure}
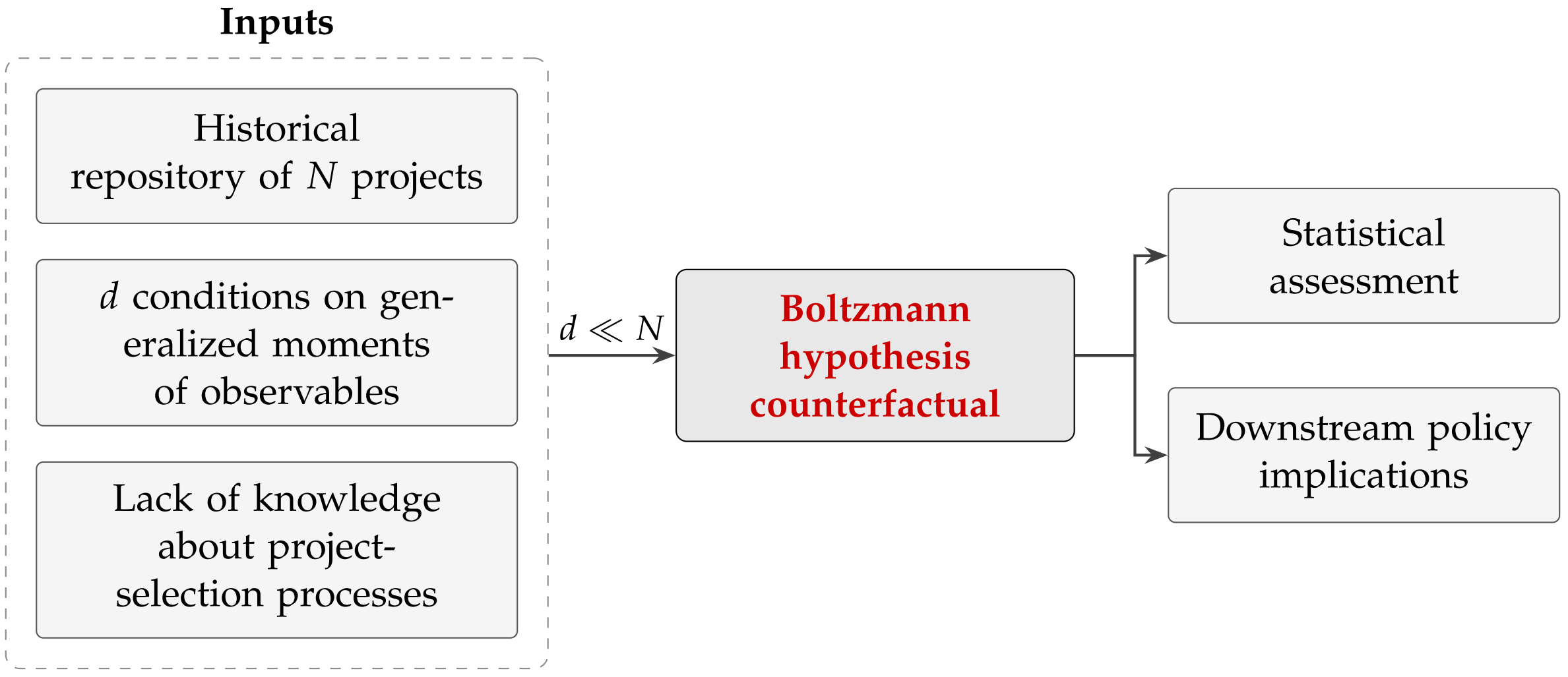

\begin{figure}[!t]
	
\begin{tikzpicture}[
	font=\sffamily,
	>={Stealth[length=2mm]},
	flow/.style={->,draw=black!65,line width=0.8pt},
	family/.style={
		draw=black!65,dashed,rounded corners=4pt,
		line width=0.8pt
	}
	]
	
	%
	\newcommand{\repository}[4]{%
		\ifnum#3=1
		\node[
		font=\scriptsize\bfseries,
		text=red!75!black
		] at (#1+1.30,#2-0.25) {Historical repository};
		\fi
		
		\foreach \n [count=\i] in {#4} {%
			\ifnum\n>0
			\foreach \j in {1,...,\n} {%
				\pgfmathsetmacro{\ballx}{#1+0.05+0.35*(\i-1)}%
				\pgfmathsetmacro{\bally}{#2+0.08+0.18*(\j-1)}%
				
				\ifcase\i
				\or \filldraw[fill=catA,draw=black!55,line width=0.25pt] (\ballx,\bally) circle (0.065);%
				\or \filldraw[fill=catB,draw=black!55,line width=0.25pt] (\ballx,\bally) circle (0.065);%
				\or \filldraw[fill=catC,draw=black!55,line width=0.25pt] (\ballx,\bally) circle (0.065);%
				\or \filldraw[fill=catD,draw=black!55,line width=0.25pt] (\ballx,\bally) circle (0.065);%
				\or \filldraw[fill=catE,draw=black!55,line width=0.25pt] (\ballx,\bally) circle (0.065);%
				\or \filldraw[fill=catF,draw=black!55,line width=0.25pt] (\ballx,\bally) circle (0.065);%
				\or \filldraw[fill=catG,draw=black!55,line width=0.25pt] (\ballx,\bally) circle (0.065);%
				\or \filldraw[fill=catH,draw=black!55,line width=0.25pt] (\ballx,\bally) circle (0.065);%
				\fi
			}%
			\fi
		}%
	}
	
	
	\draw[black!70,line width=0.8pt,rounded corners=4pt]
	(-2.0,7.55) rectangle (6.0,10.6);
	
	\node[font=\bfseries,color=catG] at (2.0,10.25)
	{Boltzmann hypothesis counterfactual};
	
	\node[font=\small,color=catG] at (2.0,9.85)
	{green-power difference $\eta\geq0$};
	
	\foreach \h/\col [count=\i] in {
		0.25/catA,
		0.45/catB,
		0.75/catC,
		1.31/catD,
		1.11/catE,
		1.33/catF,
		1.49/catG,
		1.02/catH%
	} {
		\pgfmathsetmacro{\x}{-1.1+0.85*(\i-1)}
		\fill[\col] (\x,8.1) rectangle ++(0.48,\h);
		\draw[black!55,line width=0.3pt]
		(\x,8.1) rectangle ++(0.48,\h);
	}
	
	\draw[black!65] (-1.17,8.1) -- (5.39,8.1);
	
	\node[font=\scriptsize] at (2.0,7.8)
	{historical project profiles};
	
	
	\draw[flow] (2.0,7.35) -- (-4.1,6.85);
	\draw[flow] (2.0,7.35) -- (2.0,6.85);
	\draw[flow] (2.0,7.35) -- (8.1,6.85);
	
	
	\node[font=\bfseries\small,color=catB] at (-4.1,6.5)
	{Deficit family};
	
	\node[font=\small,color=catB] at (-4.1,6.1)
	{$\mu'<\widehat{\mu}$};
	
	\repository{-5.35}{4.65}{0}{4,3,3,3,2,2,2,1} 
	\repository{-5.35}{3.20}{0}{2,4,3,3,3,1,2,2} 
	\repository{-5.35}{1.75}{0}{5,3,2,2,3,2,2,1} 
	
	\draw[family] (-5.55,1.55) rectangle (-2.45,5.8);
	
	
	\node[font=\small\bfseries,color=catC] at (2.0,6.5)
	{Historical family};
	
	\node[font=\small,color=catB] at (2.0,6.1)
	{$\widehat{\mu}<0$};
	
	\repository{0.75}{4.65}{0}{1,1,4,3,3,2,4,2} 
	\repository{0.75}{3.20}{1}{4,3,3,2,3,2,2,1} 
	\repository{0.75}{1.75}{0}{2,4,4,3,3,2,2,0} 
	
	\draw[family] (0.55,1.55) rectangle (3.65,5.8);
	
	
	
	
	\node[font=\small\bfseries,color=catF] at (8.1,6.5)
	{Surplus family};
	
	\node[font=\small,color=catF] at (8.1,6.1)
	{$\mu''>0$};
	
	\repository{6.85}{4.65}{0}{1,2,1,3,4,2,3,4} 
	\repository{6.85}{3.20}{0}{2,2,2,2,3,3,4,2} 
	\repository{6.85}{1.75}{0}{0,4,2,3,4,3,2,2} 
	
	\draw[family] (6.65,1.55) rectangle (9.75,5.8);
	
	
	
\end{tikzpicture}
\caption{\textbf{Schematic depiction of the effective  description for project selection}. 
	Given the sample size $N$, together with further structural conditions, we could --\,in principle\,-- generate all combinatorially permissible alternative repositories over historical project profiles. Based on their sample mean $\mu$, alternative project repositories are grouped  into families.  
	Under effectively independent (multinomial) selection of projects from the Boltzmann hypothesis counterfactual, which comprises project profiles in proportions satisfying the combined hypothesis of the green mandate $\eta$, alternative repositories are assigned probabilities. Coarse-graining those probabilities over the repository families gives rise to closed-form expressions to leading order in $N$. The (histogram) composition of Boltzmann hypothesis counterfactual arises as the average over alternative repositories weighted by their selection probabilities, even if no single alternative repository meets the mandated $\eta\geq0$.}\label{fig:EffectiveProjectSelection}
\end{figure}
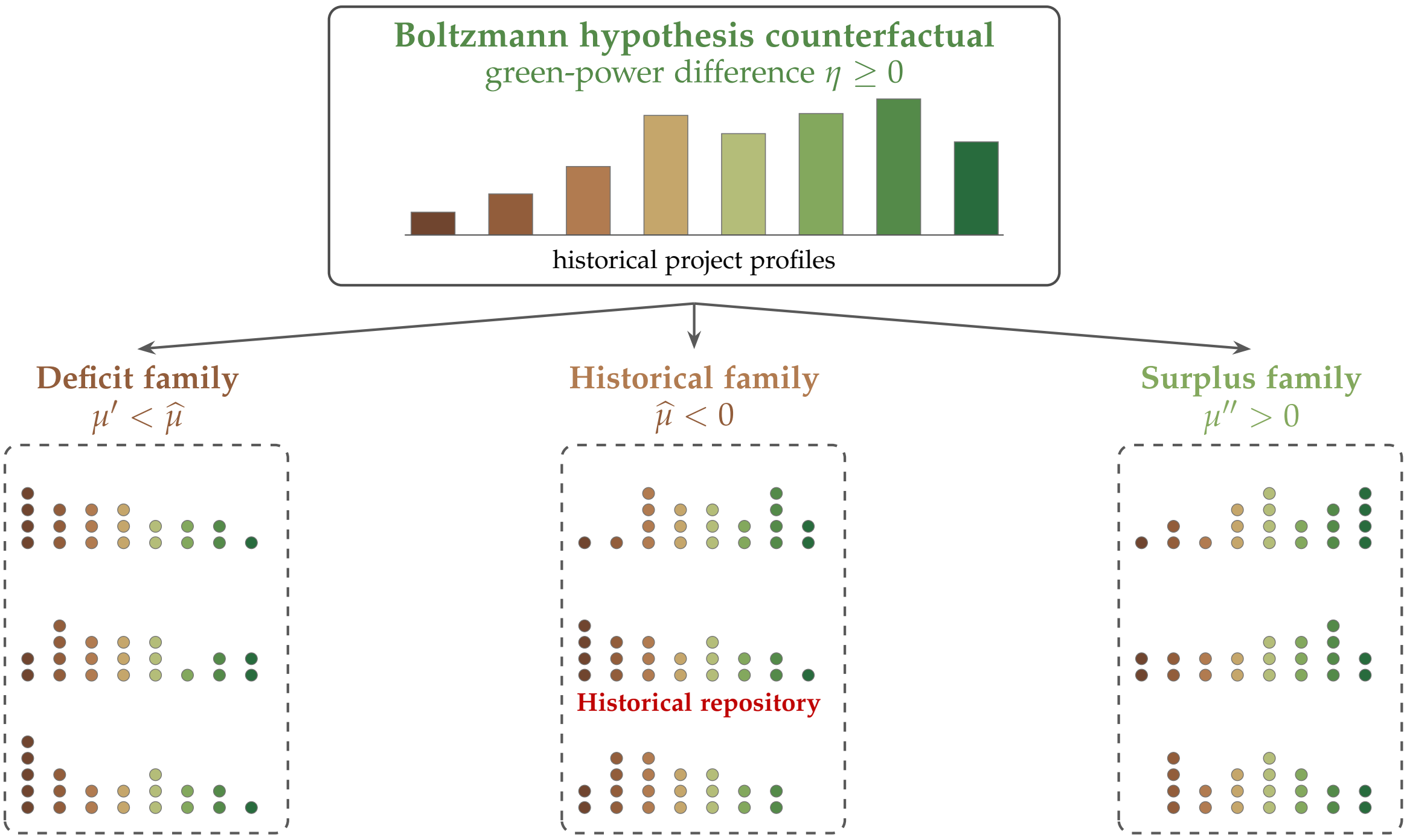


\subsection{Mathematical Background}
\label{app:Math}

Since aggregates over nameplate capacities are nothing but generalized moments of the observable $Y$, Eq.~\eqref{eq:TotalCapacityCondition} and Eq.~\eqref{eq:HydroCondition} define a structural linear family of repositories (datasets) over observed project profiles, the categories forming $\Omega$ from Eq.~(1.1) in~\cite{loukas2026boltzmannstructuresamplingintrinsic}. Nested within the structural family is the historical repository family  defined by the additional condition Eq.~\eqref{eq:GreenPowerDeficit}, which reproduces the observed green-power deficit. 
Conceptually, repository families are discrete objects determined by a system of linear Diophantine equations. 
On the simplex of Eq.~(2.21) in~\cite{loukas2026boltzmannstructuresamplingintrinsic}, we analogously define the structural $\widehat{\mathcal L}$ and nested historical $\mathcal A\equiv\mathcal L_\text{obs}$  family of distributions, Eq.~(5.2) and Eq.~(5.55) in~\cite{loukas2026boltzmannstructuresamplingintrinsic} respectively, through the $d=3$ and $D=4$ linearly independent conditions, including normalization. 

Given a hypothesis counterfactual, we take the generation of alternative repositories of size $N$, which are coarse-grained into repository families (alternative dataset families), to fall in the universality class of multinomial sampling. Such generic sampling takes place from the hypothesis distribution $\prob v=\prob h$, a member of the hypothesis linear family $\mathcal H$ of all distributions that satisfy the combined hypothesis of Section~\ref{ssc:CombinedHypothesis}. We condition the generic sampling on the structural family of repositories, which all evidence the same total and hydroelectric capacities. At large sample size $N$, the intrinsic $p$-value for the observed green-power deficit is then approximated, according to Eq.~(5.14) and Eq.~(5.69) of~\cite{loukas2026boltzmannstructuresamplingintrinsic}, by one minus the \textsc{cdf} of the $\chi^2_k$ distribution with $k=D-d=1$ degrees of freedom and test statistic given by
\begin{equation}
	\label{eq:appendix:TestStatistic}
	t = 2 N \min_{\prob p\in\mathcal A}\infdiv{\prob p}{\hat{\prob q}} \quad\text{with}\quad \hat{\prob q} = \argmin_{\prob p\in\widehat{\mathcal L}} \infdiv{\prob p}{\prob h}~. 
\end{equation}
In our current application, $\mathcal H\subset\widehat{\mathcal L}$ and thus $\hat{\prob q}=\prob h$ trivially, since $\prob h$ is itself an admissible member of $\widehat{\mathcal L}$ and hence the global minimizer of the latter information divergence. 
By construction, the historical repository is a member of the historical repository family, thus $\prob f\in\mathcal A$.

\paragraph{Derivation of the Boltzmann hypothesis distribution.}
In Section~\ref{ssc:BoltzmannDistro}, we argued for selecting a special member of the hypothesis linear family $\mathcal H$, the Boltzmann hypothesis distribution, with desirable implications both operationally and in a governance setting. We now derive the Boltzmann hypothesis distribution from a Maximum Probability condition, which makes the historical repository the mode of conditional sampling within the historical repository family under the Boltzmann hypothesis counterfactual. 

We shall first establish that, on the simplex and within Csiszár's information-geometric formulation emerging from the large-$N$ treatment of generic sampling, the Maximum Probability condition uniquely selects the Boltzmann hypothesis distribution. We then show that, also at finite $N$, the Boltzmann hypothesis distribution assigns the maximum multinomial probability to the historical repository among all repositories in the historical family. 

Consider the historical repository, whose composition is described by the empirical distribution $\prob f\in\mathcal A$. The epistemic argument for the interpretability of $p$-values selects the hypothesis distribution $\prob h\in\mathcal H$ such that $\prob f$ describes the composition of the most probable repository among those in the historical repository family.
Truncating the standard, Boltzmann-Sanov large-\(N\) expansion of the multinomial probability for sampling within the historical repository family, this requirement results in  
\begin{equation}
	\label{eq:maxProb}
	\prob f \overset{!}{=} \argmin_{\prob p \in\mathcal A}\infdiv{\prob p}{\hat{\prob q}}= \argmin_{\prob p \in\mathcal A}\infdiv{\prob p}{\prob h}~.
\end{equation}
The second equality applies because our structural conditions fix the relevant generalized moments of $Y$ by their empirical estimates. In turn, this simplifies the test statistic Eq.~\eqref{eq:appendix:TestStatistic} to $t=2N\infdiv{\prob f}{\prob h}$.

For the following derivation, we restrict attention to the case in which no observed project profile is down-weighted to zero by the combined hypothesis.
Consequently, the relevant linear families of distributions possess nonempty interiors, and all $I$-projections considered below are interior points. 
Using the above condition in Eq.~\eqref{eq:maxProb}, we can then show that  
\begin{equation}
	\label{eq:BoltzmannHypothesisDistribution}
	\prob h = \argmin_{\prob p\in\mathcal H} \infdiv{\prob p}{\prob f}~.
\end{equation}
Hence, the epistemically meaningful hypothesis distribution in the case of generic sampling under linear conditions can be understood as realizing a principle of minimal adaptation through the $I$-projection of the empirical distribution onto the hypothesis linear family.

First, we note that the hypothesis distribution would not generally factorize, like e.g.\ $h(y,x,s)=h(y,s) h(x)$, even under simple linear conditions like Eq.~\eqref{eq:TotalCapacityCondition}, since we operate over observed (and not observable) project profiles. For the remainder of the formal assertions, it is convenient to enumerate the observed profiles by index $i\in\left\lbrace1,\ldots\vert\Omega\vert\right\rbrace$ with $\vert\Omega\vert\leq N$. Because the aggregate conditions are linear in the probabilities, they can be conveniently summarized through the linear system (Eq.~(2.22) in~\cite{loukas2026boltzmannstructuresamplingintrinsic}) 
\begin{equation}
	\mathbf G\: \prob p = \boldsymbol{\mu}~.	
\end{equation} 
Here, $\boldsymbol{\mu}$ is a real-valued column-vector of generalized moments. In our application, $\hat{\boldsymbol\mu}=\left(1,\bar Y, \bar Y_\texttt{hydro}\right)^T$ for the structural family and either 
 $\boldsymbol{\mu}_A=\left(1,\bar Y, \bar Y_\texttt{hydro}, \bar Y_\texttt{green}-\bar Y_\texttt{brown}\right)^T$ for the alternative or
$\boldsymbol{\eta}=\left(1,\bar Y, \bar Y_\texttt{hydro}, 0\right)^T$ for the null-hypothesis family.  In the shorthand notation, $\bar Y$ denotes the total nameplate capacity (sample mean of $Y$), while $\bar Y_s$ denotes its aggregate over profiles with given policy-tag $s$ (and correspondingly for granular $\tilde s$).
The matrix elements $g_{a,i}$ of the appropriate coefficient matrix $\mathbf G$ (with either $d$ or $D$ rows and $\vert\Omega\vert$ columns) can be read off from the respective conditions (upon enumeration of project profiles); in our application, they are given by the nameplate capacity $Y$ and indicator functions of the policy $S$ and granular $\tilde S$ observables.

Having introduced the necessary notation and linear-algebraic formulation, we now show that Eq.~\eqref{eq:BoltzmannHypothesisDistribution} suffices for Eq.~\eqref{eq:maxProb} and, within the large-$N$ information-geometric formulation, becomes also necessary.
The sufficiency follows directly from the exponential form of the $I$-projection in Eq.~\eqref{eq:BoltzmannHypothesisDistribution}, which, for some vector of Lagrange multipliers $\widehat{\bm\theta}\in\mathbb R^{D}$, reads
\begin{equation}
	\label{eq:HypothesisDistribution:ExpoentialForm}
	h_i = f_i\: \exp\left\{\sum_{a=0}^{D-1}\widehat\theta_a\: g_{a,i}\right\} 
	~.
\end{equation}
For a feasible combined hypothesis, such a parametrization of the $I$-projection always exists, being unique up to reparametrizations of the linear system $\mathbf G\:\prob h=\boldsymbol{\eta}$ which transform the dual parameters contravariantly so that the probabilities $h_i$ remain invariant.
Substituting the form of Eq.~\eqref{eq:HypothesisDistribution:ExpoentialForm} in the leading information divergence of Eq.~\eqref{eq:maxProb} gives  
\begin{equation}
	\label{eq:KL_in_Alternative}
	\infdiv{\prob p}{\prob h} = \infdiv{\prob p}{\prob f} - \widehat{\boldsymbol\theta} \cdot {\bm\mu}_A \quad\text{for}\quad \prob p\in\mathcal A~,
\end{equation}
since all distributions $\prob p$ in the historical family reproduce the empirical estimates ${\boldsymbol{\mu}}_A$ of generalized moments. 
Because these empirical estimates  are constant for a given historical repository and the Lagrange multipliers are determined from the combined hypothesis, Eq.~\eqref{eq:KL_in_Alternative} is minimized over $\mathcal A$ whenever the information divergence from $\prob f$ is minimized. Over the historical repository family, the latter is trivially minimized at $\prob p=\prob f$, thus proving sufficiency.  

Conversely, suppose that such a $\tilde{\prob h}\in\mathcal H$ has been found according to the Maximum Probability Principle of Eq.~\eqref{eq:maxProb}. 
The mode of sampling from $\tilde{\prob h}$ conditioned on the historical repository family is captured (at sufficiently large $N$) by the $I$-projection of $\tilde{\prob h}$ onto $\mathcal A$. Rewriting its exponential form then gives,
\begin{equation}
	\tilde h_i = f_i\, \exp\left\{-\sum_{a=0}^{D-1}\widehat\lambda_a\:g_{a,i}\right\} 
	~,
\end{equation}
for some $\widehat{\bm\lambda}\in\mathbb R^{D}$.
From here, we recognize that this $\tilde{\prob h}$ is a member of the exponential family $\mathcal E$ with
\begin{equation}
	\label{eq:ExponentialFamily}
	p_i(\boldsymbol\theta)=f_i\:\exp\left(\sum_{a=0}^{D-1}\theta_a\: g_{a,i}\right)\quad\text{for}\quad\boldsymbol{\theta}\in\mathbb R^{D}~,
\end{equation}
 generated around the empirical distribution $\prob f$ by the coefficient matrix $\mathbf G$.
Consequently, $\tilde{\prob h}\in\mathcal H\cap \mathcal E$. By the duality between linear and exponential families~\cite{csiszar_information_2004}, this intersection is unique and coincides with the $I$-projection of $\prob f$ onto $\mathcal H$, thus $\tilde{\prob h}=\prob h$.

Crucially, the prescription in Eq.~\eqref{eq:BoltzmannHypothesisDistribution} for the Boltzmann hypothesis distribution continues to guarantee that the original historical repository is assigned the maximum probability in the historical repository family also at finite $N$.
In the universality class of multinomial sampling, the leading in $N$ sampling probability of any process must be dictated by the multinomial distribution (see Appendix~A of~\cite{loukas2026boltzmannstructuresamplingintrinsic}), so that it suffices to inspect 
\begin{equation}
	\label{eq:ExactMultinomialProbability}
	N\prob p\in\mathbb N_0^{\vert\Omega\vert} ~,~ \prob p\in\mathcal{A}~:\quad
	\log\operatorname{mult}(N\prob p;\prob h) = \log\operatorname{mult}(N\prob p;\prob f) + N\:\widehat{\bm\theta}\cdot{\bm{\mu}}_A \quad~,
\end{equation}
upon substituting the exponential form from Eq.~\eqref{eq:HypothesisDistribution:ExpoentialForm}. The latter term remains constant throughout the alternative family, since it only depends on the original composition $N\prob f$, the coefficient matrix $\mathbf G$ and vector of hypothesized generalized moments $\boldsymbol{\eta}$. 

Consequently, the exact maximization of the multinomial probability from $\prob h$ over the historical repository family reduces to the corresponding maximization from $\prob f$. 
It is well known that the multinomial distribution with parameters (intrinsic probabilities) $f_i$ is globally maximized at the count vector $N\prob f$, which contains the historical profile frequencies. Since the historical repository family (as any other repository family) is a subset of the possible repositories of size $N$ and contains by definition the historical repository itself, the count vector $N\prob f$ necessarily maximizes the multinomial probability over the historical repository family. By the preceding decomposition in Eq.~\eqref{eq:ExactMultinomialProbability}, it therefore also maximizes the sampling probability from $\prob h$. This finite-\(N\) sufficiency argument  holds for any \(\vert\Omega\vert\leq N\).

\paragraph{Attributability of counterfactual generalized moments to historical baseline.}
Following the information-theoretic derivation of $\prob h$ and after asserting its epistemically desirable property, we now expose a more stringent property of the Boltzmann hypothesis distribution that renders it unique among the members of $\mathcal H$. In the Boltzmann hypothesis counterfactual, the downstream value of    any generalized moment left unspecified by the combined hypothesis of Section~\ref{ssc:CombinedHypothesis} is uniquely determined through attributability under conditional sampling. Specifically, the conditional sample estimate\footnote{Evidently, the unconditional sample estimate will fluctuate around the counterfactual value prescribed by the hypothesis distribution.} of such a generalized moment fluctuates around its historical value only if the alternative repositories supporting the historical green-power deficit are generated under the Boltzmann hypothesis counterfactual. 

To show this when both $N$ and the number of project profiles exceed the number of imposed conditions, we consider a function $\phi:\Omega\rightarrow\mathbb R$ which prescribes a generalized moment that may ultimately enter the definition of some downstream metric. We are interested in otherwise arbitrary generalized moments that are not implied by the $Y$-moments entering the structural and mandated conditions. 
Within the structural repository family, we then construct two alternative repository families that both adhere to the structural conditions.  The historical repository family is defined  by specifying the observed power deficit Eq.~\eqref{eq:GreenPowerDeficit}, as before. A nested repository family additionally supports the historical estimate for the generalized moment,
\begin{equation}
	\label{eq:ArbitraryTestFeature}
	\sum_{i=1} p_i\, \phi_i\overset{!}{=} \sum_{i=1} f_i\, \phi_i~,
\end{equation}
hence in total defined by five linearly independent conditions, including normalization.

Given some hypothesis distribution $\tilde{\prob h}\in\mathcal H$, we perform a nested hypothesis test that captures the statistical plausibility of the historical estimate of the generalized moment given the observed green-power deficit. Formally, we compare the ratio of sampling probabilities of the nested to the historical family. A ratio value close to one would signify that the additional condition in Eq.~\eqref{eq:ArbitraryTestFeature} entails little loss of sampling probability mass, so that deviations from the historical sample mean of \(\phi\) could be attributed to non-systematic fluctuations within the historical repository family.
Hence, we would conclude that knowledge of the observed green-power deficit would statistically imply the historical estimate of the additional generalized moment  under the specified structural conditions. 

Saddle-point approximation to this ratio helps compute the intrinsic $p$-value to observe the historical estimate or any at least as improbable value of the generalized moment, given the historical green-power deficit. The test statistic of approximating $\chi^2_1$ is given by 
\begin{equation}
	\label{eq:Nested_pValue}
	t = 2 N \min_{\prob p \in\mathcal A_\text{nested}}\infdiv{\prob p}{\prob q} \quad\text{with}\quad \prob q = \argmin_{\prob p\in\mathcal A} \infdiv{\prob p}{\tilde{\prob h}}~,
\end{equation}
where the historical family $\mathcal A$ plays now the role of the ambient linear family on the simplex, cf.\ Eq.~\eqref{eq:appendix:TestStatistic}. 
As shown above, $\prob q$, being the $I$-projection onto the historical linear family, will not coincide with $\prob f$, unless we select the Boltzmann hypothesis distribution. Similarly, $\prob p = \prob q$ does not need to fulfill Eq.~\eqref{eq:ArbitraryTestFeature} and thus the ambient $I$-projection will not generally belong to the nested family $\mathcal A_\text{nested}\subset\mathcal A$.

Requiring the $\tilde{\prob h}$-counterfactual structure that is captured  by any such generalized moment to be fully attributable to the historical repository, given the historical green-power deficit, translates to an intrinsic $p$-value of one in Eq.~\eqref{eq:Nested_pValue}.
Colloquially, we demand that knowledge of the green-power deficit implies the historical estimate of any other generalized moment in the universe of the selected hypothesis counterfactual. The Boltzmann hypothesis counterfactual automatically fulfills this requirement, since $\prob q=\prob f\in\mathcal A_\text{nested}$ for $\tilde{\prob h}=\prob h$ and thus $t=0$.

Let us now select another $\tilde{\prob h}\neq\prob h$ which yields $\prob q\neq\prob f$. Since \(\phi\) is arbitrary and \(\Omega\) is finite, whenever \(\prob q\neq\prob f\), we can always choose a generalized moment so that the $I$-projection onto the ambient historical family does not verify Eq.~\eqref{eq:ArbitraryTestFeature} and hence $\prob q\not\in\mathcal A_\text{nested}$.  In turn, this gives a non-trivial $I$-projection onto the nested family, meaning $t>0$ in Eq.~\eqref{eq:Nested_pValue}. In total, choosing the Boltzmann hypothesis distribution is necessary and sufficient to ensure attributability to the historical baseline of the counterfactual estimate for any individually considered generalized moment. The argument generalizes to nested testing of multiple generalized moments given the observed green-power deficit.

 \paragraph{On the explicit dependence of the intrinsic test statistic on the empirical distribution.}
 Equation~\eqref{eq:maxProb} has a further interpretational consequence that implies a desirable property for the intrinsic test statistic. Another hypothesis distribution $\tilde{\prob h}$, selected through some alternative mechanism, would generically lead, after coarse-graining sampling probabilities over repository families, to a test statistic for the intrinsic $p$-value of the form
 $t'=2N\infdiv{\tilde{\prob q}}{\tilde{\prob h}}$,
 where $\tilde{\prob q}\neq\prob f$ is the $I$-projection of $\tilde{\prob h}$ onto the historical family of distributions. 
 
 Suppose now that the selection mechanism for $\tilde{\prob h}$ is blind to a change in the empirical distribution $\prob f$, which leaves --\,besides $\tilde{\prob h}$\,-- the generalized moments defining the structural and hypothesis families invariant. For the underspecified combined hypothesis considered in the policy setting, such directions always exist. Consequently, the $I$-projection $\tilde{\prob q}$ and the resulting statistic $t'$ remain invariant. In other words, the empirical information contained in such changes has been lost, effectively replaced by the ``model''-dependent prescription defining $\tilde{\prob q}$ through the selection mechanism for $\tilde{\prob h}$. 
 
 By contrast, Eq.~\eqref{eq:maxProb} ensures that the test statistic of the intrinsic $p$-value reduces to
 $t=2N\infdiv{\prob f}{\prob h}$,
 so that changes in the historical repository enter the test statistic directly rather than being absorbed after coarse-graining. In this sense, the Maximum Probability condition prevents the intrinsic $p$-value from acquiring an indirect, ``model''-dependent test statistic. Within our Boltzmann construction, if certain aspects of the repository's composition are to be adjusted, these can be specified directly as additional structural conditions, whose generalized moments need not necessarily be given by their empirical estimates.
 
\paragraph{Relation to model-centred literature.}
The exponential-tilting form in Eq.~\eqref{eq:HypothesisDistribution:ExpoentialForm} of the hypothesis distribution derived here bears formal resemblance to tilted constructions in econometrics. In the literature, the information divergence (among other divergences) is minimized to construct probability weights satisfying moment restrictions for estimation within an assumed statistical model, in accordance with the axiomatic formulation of probability theory; see, e.g.~\cite{kitamura1997information,imbens1998information,schennach2007pointestimation}. 

In our non-axiomatic construction, by contrast, the exponential form emerges from large-$N$ generic sampling in conjunction with the Maximum Probability condition. This selects the Boltzmann hypothesis distribution by requiring the historical repository to be the most probable member of the historical repository family. The Boltzmann hypothesis distribution then enters the coarse-graining of sampling probabilities over repository families, from which the interpretable intrinsic $p$-value is derived.

\paragraph{Typicality of the historical repository.}
Interestingly, the Principle of Maximum Probability in Eq.~\eqref{eq:maxProb} determines (to $N$-leading order) the expected frequencies when sampling over the historical repository family, at least\footnote{Since $\boldsymbol{\delta}\in\ker\mathbf G\Rightarrow-\boldsymbol{\delta}\in\ker\mathbf G$, higher-order cancellations among fluctuations around $N\prob f$ could ensure the stability of estimate Eq.~\eqref{eq:f_as_expectedDataset} even when a ``naive'' Gaussianization in the continuum limit fails under $\vert\Omega\vert=\order{N}$.} as long as $\vert\Omega\vert \ll N$ within the universality class of multinomial sampling. Generalizing the usual entropy-concentration arguments~\cite{jaynes_concentration_1989}, we may estimate in Laplace approximation 
\begin{equation}
	\label{eq:f_as_expectedDataset}
	\sum_{\mathbf n\in\mathbb N_0^{\vert\Omega\vert}}\mathbf n\operatorname{Pr}(\mathbf n\vert\mathrm A; \prob h) = N \prob f + \order{1}~,
\end{equation}
using the Gaussianization of the sampling probability density in the continuum limit around the $I$-projection of $\prob h$ onto $\mathcal A$, see e.g.\ section~2 of~\cite{loukas2026boltzmannstructuresamplingintrinsic}. The summation goes over all repositories in the historical repository family $\mathrm A$, i.e.\ over all count vectors $\mathbf n$ that reproduce the observed rescaled generalized moments  $N\bm\mu_A=\mathbf G\:\mathbf n$. 

Hence, the epistemic argument 
makes the historical repository not only the most probable, but also \textit{typical} in expectation under conditional sampling from the Boltzmann hypothesis distribution. 
In other words, under the Boltzmann hypothesis counterfactual, the historical relative frequencies maximize the sampling probability and differ from the corresponding expected relative frequencies by corrections of order $1/N$, 
when sampling alternative repositories in the historical repository family.

\subsection{Further downstream analysis}
\label{app:DownstreamAnalysis}

We complement the downstream analysis of Section~\ref{ssc:DownstreamAnalysis} by analysing further potentially interesting dimensions of the Boltzmann hypothesis counterfactuals constructed for the green transition.

\paragraph{Reweighting across nameplate capacities.}
\begin{figure}[t]
	\centering
	\includegraphics[width=1.\linewidth]{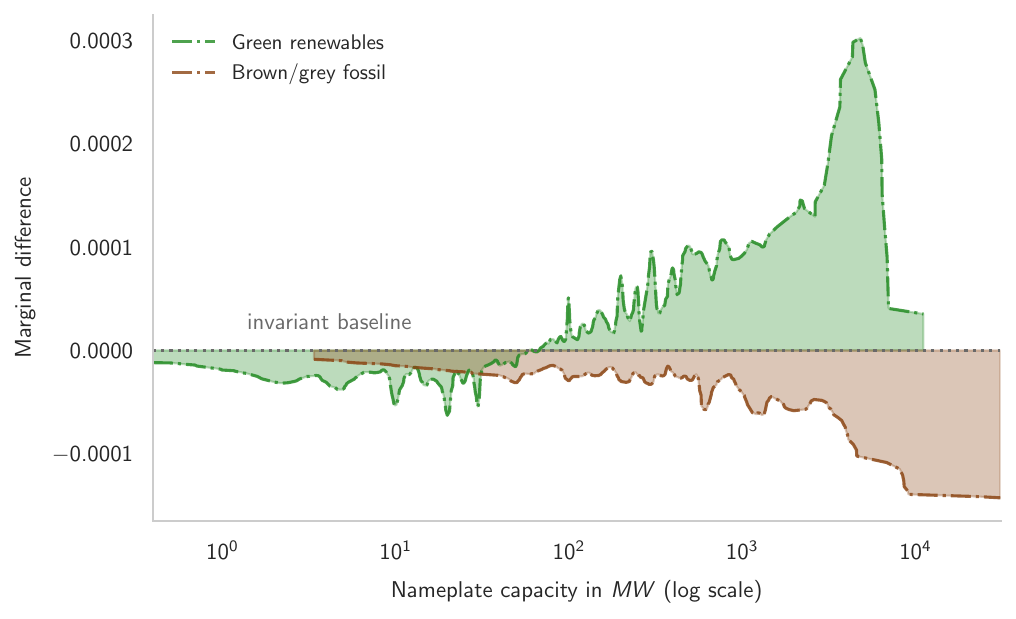}
	\caption{\textbf{Reweighting across nameplate capacities under the null hypothesis}. The curves depict the difference between the marginal distribution of $Y$ and $S$ in the hypothesis counterfactual of null power difference and in the historical repository, shown separately for each technology tag. For clarity, the curves are smoothed with a Gaussian filter of width $\sigma=5$ to reveal how the counterfactual reweighting varies over the complete range of observed nameplate capacities at exact green-brown power balance. Before smoothing, the probability mass added above the zero-difference baseline is exactly balanced by the mass removed below it, preserving total probability normalization.
	}
	\label{fig:TiltingWeights}
\end{figure}

To examine more closely the reweighting of project profiles underlying the mandated transition, we consider the marginal distribution of nameplate capacity $Y$ and policy observable $S$. In Figure~\ref{fig:TiltingWeights}, we compare the historical marginal distribution $f(y,s)$ with the corresponding marginal of the Boltzmann hypothesis distribution $h_\text{null}(y,s)$ at the null power difference, where total green and brown power are exactly balanced. 

The reweighting shown in Figure~\ref{fig:TiltingWeights} reveals a deeper restructuring than a simple shift in the prevalence of the two technology tags. In particular, green project profiles with higher nameplate capacities are up-weighted, while green profiles at the lower-$Y$ end are down-weighted.  At the same time, brown profiles are down-weighted, with the reduction becoming more pronounced toward the higher-$Y$ end. The fixed aggregate hydroelectric capacity, which constrains green profiles at the highest-$Y$ end, further limits the extent to which additional green power can be gained through such hydroelectric profiles.
Hence, the resulting reweighting scheme depends non-trivially on the technology tag and the nameplate capacity of the project profile.\footnote{In particular, our combined hypothesis does not make the ansatz $h_\text{null}(y,s)=f(y\mid s)h_\text{null}(s)$ that constrains the conditional distribution of nameplate capacity given the technology tag to equal its historical counterpart, i.e.\ we do not contemplate the more restrictive set of conditions $h_\text{null}(y\mid s)=f(y\mid s)$ for all observed tuples $(y,s)$.} 

As clearly illustrated, satisfying a small number of aggregate conditions induces a non-trivial redistribution of probability mass across individual project profiles that is neither immediately apparent from the aggregate conditions themselves nor  tractable in closed form. Generically, the effect of reweighting propagates non-trivially across all dimensions considered in the downstream analysis through the non-trivial associations encoded in the empirical distribution, even when the combined hypothesis constrains only the $Y$ and $S$ observables. 

\paragraph{Stability of private ownership participation.}
\begin{figure}[!t]
	\centering
	\includegraphics[width=1\linewidth]{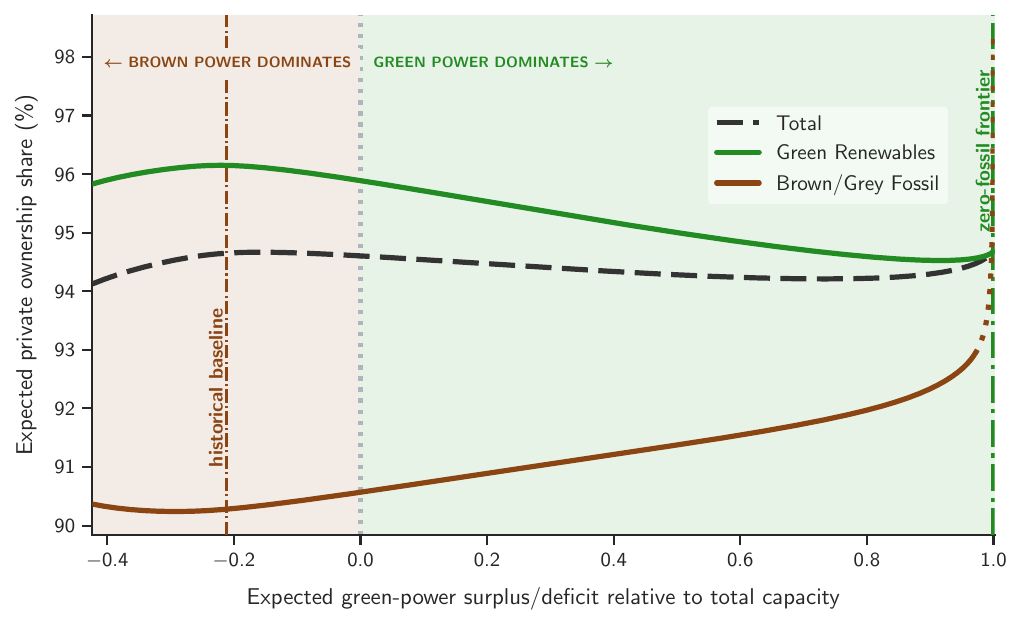}
	\caption{\textbf{Expected private ownership share}. Conditional means of private ownership share stratified by technology tag and total mean across the Boltzmann hypothesis counterfactuals.}
	\label{fig:InvestmentShare}
\end{figure}
Having analyzed the counterfactual joint statistics of $Y$ and $X$ with $S$, we now turn to the secondary observable $Z_1$, corresponding to the \texttt{PercentPrivate} attribute, which records the percentage of the project company owned by private sponsors~\cite{WorldBankPPI_Glossary}. 
Again, we stratify the downstream analysis by the policy observable $S$, i.e.\ we consider the conditional marginal probabilities $h(z_1\vert s)$ alongside the marginal probabilities $h(z_1)$ in the hypothesis counterfactual.

In Figure~\ref{fig:InvestmentShare}, we examine the expected private ownership share across the targeted power differences. The conditional means stratified by technology tag reveal a pretty stable private-sector ownership participation throughout the transition toward green-power dominance. Around the historical baseline, the expected private ownership share within green projects is slightly above $96\%$. As the mandated power difference increases toward green dominance, this share declines gradually but remains above $94\%$ even at the zero-fossil frontier.

The corresponding share for brown projects remains systematically lower, fluctuating within a relatively narrow range between $90\%$ and $93\%$ across the region in which brown projects remain prevalent. The apparent sharp increase (signified by the brown dotted segment) as the zero-fossil frontier is approached does not reflect an instability of the Boltzmann reweighting. Rather, it results from the vanishing prevalence of brown projects, $h(\texttt{brown})\to0$. As the prevalence of brown projects becomes progressively smaller, the corresponding conditional mean becomes increasingly sensitive and ultimately undefined at the zero-fossil limit.

The overall expected private ownership share also varies only slightly, remaining above $94\%$. Recall that the total mean is the prevalence-weighted average of the two conditional means, with the tagged prevalences $h(s)$ themselves varying along the targeted power differences (viz.\ Figure~\ref{fig:Prevalence}). Thus, within the hypothesis counterfactuals, aggregate private ownership participation remains largely insensitive to the targeted environmental mandate. In particular, the counterfactual does not indicate a substantial displacement of private ownership by public ownership as the targeted power difference moves toward green dominance.

\paragraph{Mild temporal restructuring of the repository.}
\begin{figure}
	\centering
	\includegraphics[width=1.\linewidth]{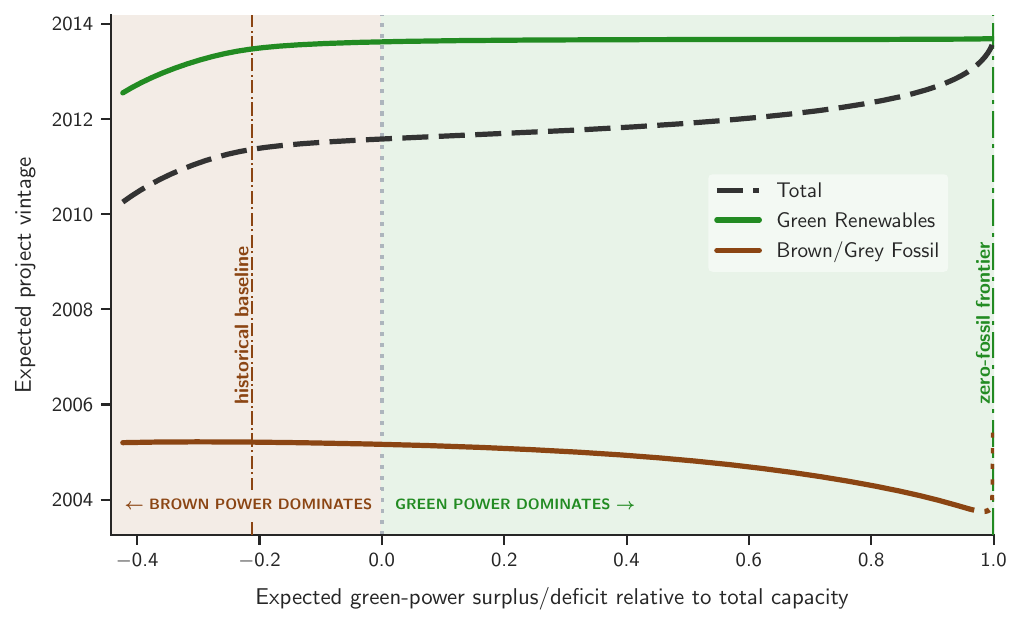}
	\caption{\textbf{Expected project vintage}. Conditional means of project vintage stratified by tagged technology and the total mean across the Boltzmann hypothesis counterfactuals.}
	\label{fig:Vintage}
\end{figure}
Having examined the fiscal and ownership consequences of the green transition, we now turn to the temporal composition of the historical repository, which contains electricity-generation projects over the 40-year period from 1984 to 2024. 
To analyze how this composition adapts to achieve the green mandate in the Boltzmann-hypothesis counterfactual, we examine in Figure~\ref{fig:Vintage} the expected project vintage across targeted power differences. The observable $T$, representing investment year, belongs to the economic group of core observables introduced for bookkeeping in Section~\ref{ssc:CompleteProjectProfiles}. Again, we stratify the downstream analysis by the policy observable $S$, i.e.\ we consider the conditional marginal probabilities $h(t\vert s)$ alongside the marginal probabilities $h(t)$ in the hypothesis counterfactual.

The tagged conditional means of project vintage remain relatively stable across the transition. The green trajectory increases gradually from the historical baseline and stabilizes shortly below 2014, already within the brown-dominance region, while the brown trajectory declines only marginally from slightly above 2005 to approximately 2004. The total mean vintage therefore shifts from approximately 2010 at the extreme deficit to 2011 at the historical baseline and subsequently approaches the green conditional mean as the prevalence of brown projects progressively vanishes.
The known numerical instabilities of the brown conditional mean as its prevalence vanishes toward the zero-fossil frontier are depicted with the same dotted-line convention used above.

\paragraph{Geographic allocation.}

\begin{figure}[t!]
	\centering
	\includegraphics[width=1.01\linewidth]{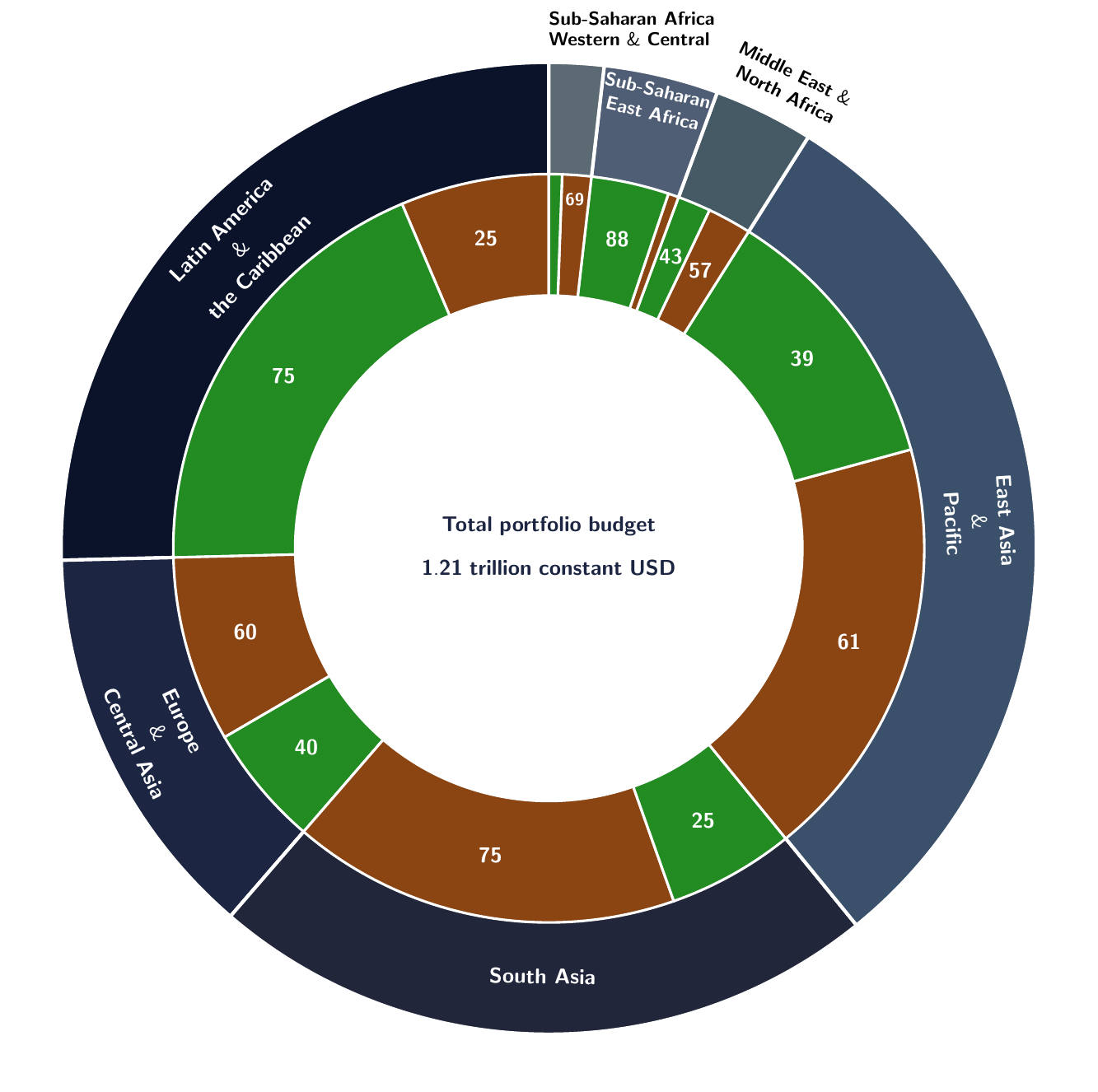}
	\caption{\textbf{Regional investment shares in the historical repository}. The concentric pie chart shows the allocation of total investment across the seven macro-regions, with each regional share further partitioned between green and brown projects. Percentages shown in the inner ring are rounded to the nearest integer; the \% symbol is omitted for visual clarity.}
	\label{fig:budget_regions1}
\end{figure}
\begin{figure}[t!]
	\centering
	\includegraphics[width=1.01\linewidth]{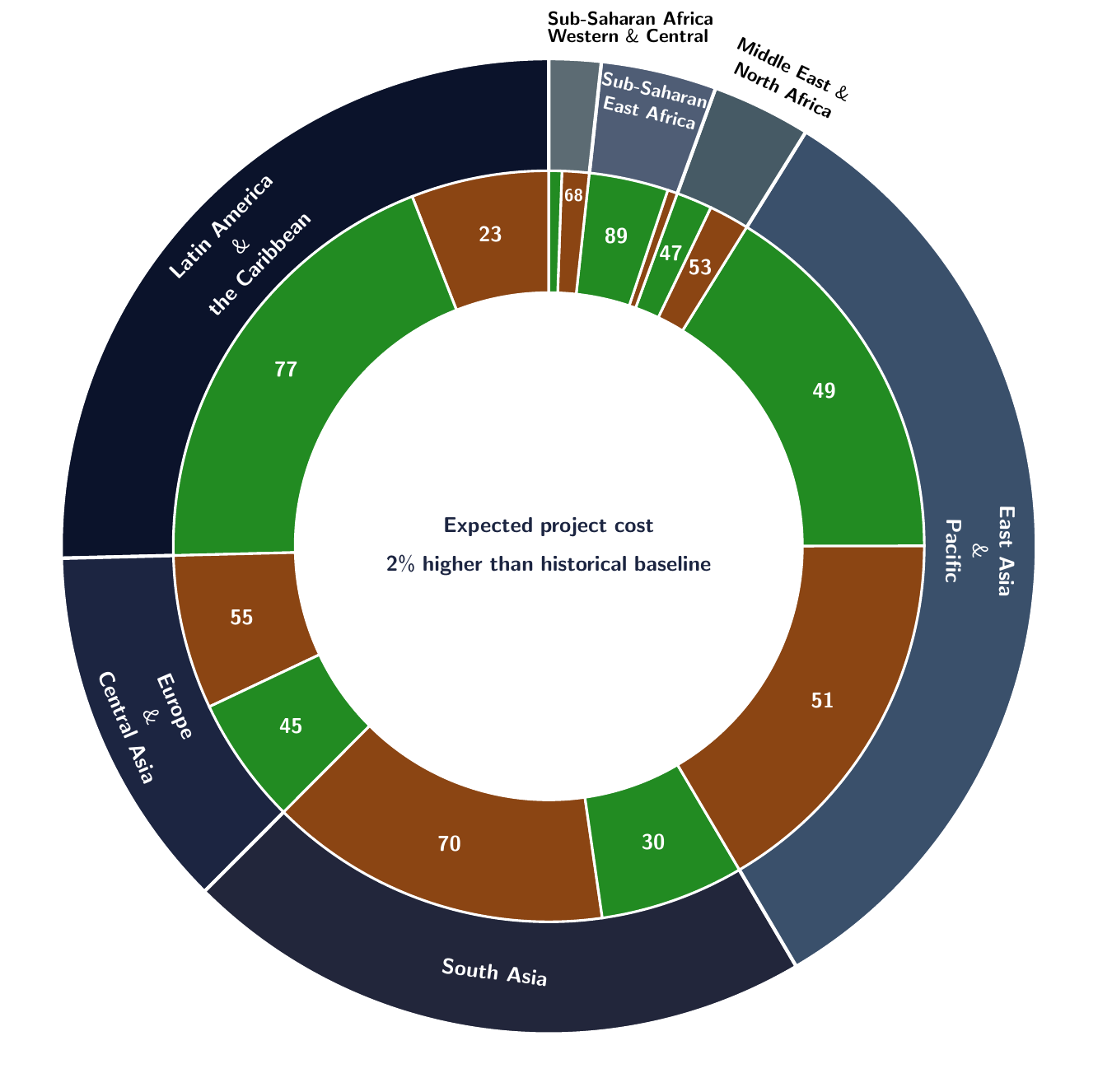}
	\caption{\textbf{Regional investment shares under green-brown power parity, $\eta=0$ in Eq.~\eqref{eq:EtaCondition}}.}
	\label{fig:budget_regions2}
\end{figure}

We conclude the downstream analysis by examining the spatial dimension in the constructed counterfactuals through the joint marginal of the fiscal observable \(X\), policy observable \(S\), and secondary observable \(Z_2\) corresponding to the \texttt{Region} attribute~\cite{WorldBankPPI_Glossary}. The fiscal (re)allocation is visualized through concentric pie charts in Figures~\ref{fig:budget_regions1} and~\ref{fig:budget_regions2}. The outer ring displays the investment share in each region,
\begin{equation}
	\sum_x x p(x,z_2)\bigg / \sum_x x p(x) \quad\text{for}\quad z_2\in\{\texttt{Latin\,\,America\,\,\&\,\,the\,\,Caribbean},\ldots\}~,
\end{equation} 
while the inner ring partitions each regional share into green and brown investment, 
\begin{equation}
	\sum_x x p(x,z_2,s)\bigg / \sum_x x p(x,z_2) ~.
\end{equation} 

Figure~\ref{fig:budget_regions1} charts the historical allocation ($\prob p=\prob f$) of total investment across the seven geographic regions in the World Bank database, explicitly stratifying what proportion of each region's budget is allocated to green versus brown project profiles. Figure~\ref{fig:budget_regions2} charts the spatial composition in the Boltzmann hypothesis counterfactual ($\prob p=\prob h_\text{null}$) of the transition regime at null power difference.
A direct comparison of these two spatial compositions yields a by now anticipated result. The investment share allocated to each macro-region remains virtually unaltered, displaying barely visible adjustments, since the green mandate is mainly satisfied through localized internal shifts from brown to green investment within each macro-region.

\end{document}